\documentclass[twocolumn]{aastex6}

\usepackage{tabularx}
\usepackage{epsfig}
\usepackage[english]{babel}
\usepackage[utf8]{inputenc}
\usepackage{amsmath,amsfonts}
\usepackage{url}

\usepackage{comment}
\usepackage{multirow}

\usepackage[titletoc]{appendix}
\usepackage{natbib}
\usepackage{ulem}

\def \aap{A\&A}
\def \apjl{ApJ}

\def \apjs{ApJS}

\begin{document}

\title{The Galah Survey: Classification and diagnostics with t--SNE reduction of spectral information}

\author{
G.~Traven \altaffilmark{1},
G.~Matijevič \altaffilmark{2},
T.~Zwitter \altaffilmark{1},
M.~Žerjal \altaffilmark{1},
J.~Kos \altaffilmark{3},
M.~Asplund \altaffilmark{4},
J.~Bland-Hawthorn \altaffilmark{3},
A.~R.~Casey \altaffilmark{5},
G.~De~Silva \altaffilmark{6,3},
K.~Freeman \altaffilmark{4},
J.~Lin \altaffilmark{4},
S.~L.~Martell \altaffilmark{7},
K.~J.~Schlesinger \altaffilmark{8},
S.~Sharma \altaffilmark{3},
J.~D.~Simpson \altaffilmark{9,10},
D.~B.~Zucker \altaffilmark{10,11,6},
B.~Anguiano \altaffilmark{10},
G.~Da~Costa \altaffilmark{4},
L.~Duong \altaffilmark{4},
J.~Horner \altaffilmark{12},
E.~A.~Hyde \altaffilmark{13,14},
P.~R.~Kafle \altaffilmark{15},
U.~Munari \altaffilmark{16},
D.~Nataf \altaffilmark{4,17},
C.~A.~Navin \altaffilmark{18,10},
W.~Reid \altaffilmark{18,19},
Y.--S.~Ting \altaffilmark{20}
}

\altaffiltext{1}{Faculty of Mathematics and Physics, University of Ljubljana, Jadranska 19, 1000 Ljubljana, Slovenia; \\ e-mail: gregor.traven@fmf.uni-lj.si}
\altaffiltext{2}{Leibniz-Institut f\"ur Astrophysik Potsdam (AIP), An der Sternwarte 16, 14482 Potsdam, Germany}
\altaffiltext{3}{Sydney Institute for Astronomy, School of Physics, A28, The University of Sydney, NSW 2006, Australia}
\altaffiltext{4}{Research School of Astronomy and Astrophysics, Australian National University, Canberra, ACT 2611, Australia}
\altaffiltext{5}{Institute of Astronomy, University of Cambridge, Madingley Road, Cambridge CB3 0HA, UK}
\altaffiltext{6}{Australian Astronomical Observatory, PO Box 915, North Ryde, NSW 1670, Australia}
\altaffiltext{7}{School of Physics, University of New South Wales, Sydney, NSW 2052, Australia}
\altaffiltext{8}{Research School of Astronomy \& Astrophysics, Mount Stromlo Observatory, Cotter Road, Weston Creek, ACT 2611, Australia}
\altaffiltext{9}{Australian Astronomical Observatory, North Ryde, NSW 2113, Australia}
\altaffiltext{10}{Department of Physics and Astronomy, Macquarie University, North Ryde, NSW 2109, Australia}
\altaffiltext{11}{Research Centre in Astronomy, Astrophysics \& Astrophotonics, Macquarie University, North Ryde, NSW 2109, Australia}
\altaffiltext{12}{Computational Engineering and Science Research Centre, University of Southern Queensland, Towoomba QLD 4350, Australia}
\altaffiltext{13}{Western Sydney University, Locked Bag 1797, Penrith, NSW 2751, Australia}
\altaffiltext{14}{Australian Astronomical Observatory, PO Box 296 Epping, NSW 1710, Australia}
\altaffiltext{15}{ICRAR, The University of Western Australia,35 Stirling Highway, Crawley, WA 6009, Australia}
\altaffiltext{16}{INAF National Institute of Astrophysics, Astronomical Observatory of Padova, 36012 Asiago (VI), Italy}
\altaffiltext{17}{Department of Physics and Astronomy, Johns Hopkins University, Baltimore, MD, USA}
\altaffiltext{18}{Department of Physics and Astronomy, Macquarie University, Sydney, NSW 2109, Australia}
\altaffiltext{19}{Western Sydney University, Penrith South DC, NSW 1797}
\altaffiltext{20}{Harvard Center for Astrophysics. 60 Garden Street, Cambridge, MA 02138, USA}

\begin{abstract}
Galah is an ongoing high-resolution spectroscopic survey with the goal of disentangling the formation history of the Milky Way, using the fossil remnants of disrupted star formation sites which are now dispersed around the Galaxy. It is targeting a randomly selected, magnitude limited ($V \leq 14$) sample of stars, with the goal of observing one million objects. To date, 300,000 spectra have been obtained. Not all of them are correctly processed by parameter estimation pipelines and we need to know about them. We present a semi-automated classification scheme which identifies different types of peculiar spectral morphologies, in an effort to discover and flag potentially problematic spectra and thus help to preserve the integrity of the survey's results. To this end we employ a recently developed dimensionality reduction technique t--SNE (t--distributed Stochastic Neighbour Embedding), which enables us to represent the complex spectral morphology in a two--dimensional projection map while still preserving the properties of the local neighbourhoods of spectra. We find that the majority (178,483) of the 209,533 Galah spectra considered in this study represents normal single stars, whereas 31,050 peculiar and problematic spectra with very diverse spectral features pertaining to 28,579 stars are distributed into 10 classification categories: \textit{Hot stars}, \textit{Cool metal-poor giants}, \textit{Molecular absorption bands}, \textit{Binary stars}, \textit{H$\alpha$/H$\beta$ emission}, \textit{H$\alpha$/H$\beta$ emission superimposed on absorption}, \textit{H$\alpha$/H$\beta$ P-Cygni}, \textit{H$\alpha$/H$\beta$ inverted P-Cygni}, \textit{Lithium absorption}, and \textit{Problematic}. Classified spectra with supplementary information are presented in the catalogue, indicating candidates for follow-up observations and population studies of the short-lived phases of stellar evolution.
\end{abstract}

\maketitle

\section{Introduction}

In recent decades, the technology of optical fibre--fed spectrographs has enabled very efficient large--scale automated spectroscopic surveys. With the ability to observe up to several hundred stars simultaneously, it is now possible to obtain large numbers of high quality spectra in a reasonable amount of time \citep{1987PhDT.........7W}. Surveys such as the RAVE (RAdial Velocity Experiment;  \citealp{2006AJ....132.1645S}), the APOGEE (Apache Point Observatory Galactic Evolution Experiment; \citealp{2015arXiv150905420M}), the LAMOST survey (Large Sky Area Multi-Object Fiber Spectroscopic Telescope; \citealp{2015RAA....15.1095L}), the ongoing Gaia-ESO Survey \citep{2012Msngr.147...25G}, the Galah (GALactic Archaeology with Hermes; \citealp{2015MNRAS.449.2604D}), and the Gaia mission \citep{2012AN....333..453P} with its future follow-up projects WEAVE \citep{2012SPIE.8446E..0PD} and 4MOST (4-metre Multi-Object Spectroscopic Telescope; \citealp{2012SPIE.8446E..0TD}) are some of the leading examples of continuous production of overwhelming amounts of data.  

To get a general overview of the observed spectra and learn more about the studied sample of stars in a spectroscopic survey such as Galah, it seems reasonable and necessary to address this task in an unbiased and automated way. A common approach is to employ different numerical dimensionality reduction methods to reveal the complex morphological structure of the dataset at hand. By projecting the spectra into a low dimensional space, it becomes feasible to grasp their inter-correlations and identify diverse morphological groups, thus constructing a classification of the whole dataset, and particularly its outstanding features. A plethora of linear and non-linear mathematical techniques has been developed in the past decades to tackle the problem of classification of complicated high--dimensional data such as spectra, and they have been successfully applied in the astronomical community as well. 

Many authors have used different techniques in order to classify stellar and other types of spectra (galaxy, quasar) or to discover new classes and unusual objects: \cite{1994ApJ...426..340G} and \cite{1994MNRAS.269...97V} were among the first ones using artificial neural networks; \cite{1997AJ....113.1865I}, \cite{1998MNRAS.298..361B}, and \cite{2010AJ....139.1261M} demonstrated the use of PCA (Principal Component Analysis) as a very robust classifier; \cite{2011AJ....142..203D} and \cite{2012ApJS..200...14M} succesfully used the LLE method (Locally Linear Embedding) to identify anomalous or peculiar spectra as well as classify normal ones. Apart from the last two cases, where the authors used the non-linear dimensionality reduction technique LLE, these studies have mostly dealt with discrimination between classes of the MKK scheme. However, many have mentioned the potential to identify and characterise spectral populations, which is the focus of the present work.

Galah is an ongoing spectroscopic survey that aims to unveil the Milky Way's history by studying the fossil record of ancient star formation and accretion events preserved in stellar light. Detailed knowledge of the chemical information of fossil remnants, which have disrupted and are now dispersed around the Galaxy, is essential to disentangle its formation history and explain its current stellar populations. Recent studies of chemical abundances of stars in individual (undispersed) open clusters show that their abundance distributions are homogeneous to the level at which they can be measured, and their abundances are different from cluster to cluster (e.g. \citealp{2007AJ....133.1161D}; conversely, most globulars show inhomogenities, e.g. Na$-$O anti-correlation, e.g. \citealp{2009AA...505..117C,2012AARv..20...50G}; furthermore, small abundances variations have been detected in star-to-star studies in open clusters, e.g. \citealp{2016MNRAS.457.3934L,2016MNRAS.tmp.1165L}). This enables the technique of chemical tagging \citep{2002ARAA..40..487F} to identify the fossil remnants of old dispersed clusters from their abundance patterns over many chemical elements. Galah will achieve this by measuring up to 29 elemental abundances from 7 independent element groups each with 5 measurable abundance levels, thereby obtaining enough cells ($5^7$) in multi--dimensional chemical abundance space (C-space), in which stars from chemically homogeneous aggregates (e.g. disrupted open clusters) will lie in tight clumps \citep{2012ASPC..458..393F,2012MNRAS.421.1231T}. This level of accuracy and the amount of elemental abundance information by far surpasses any existing single or multiple system stellar studies. 

The Galah automatic pipeline is currently running without a classification processing stage. By manually scanning the observed sample, it has become obvious that there is a significant number of peculiar and otherwise problematic spectra. Although the majority belong to single stars and can be properly fit by synthetic spectra, neglecting the outliers can lead to erroneous results in radial velocities, atmospheric parameters, and especially detailed chemical abundances. Finding outliers by comparison to databases of known peculiar spectra might produce useful results, but would fail to give a reliable classification of the whole sample. We therefore aim to diagnose and classify the diverse morphologies in the Galah dataset, with the goal of: (1) highlighting all problematic spectra with unpredictable effects from either instrumentation or reduction stages, (2) identifying any peculiar spectra that are interesting per se and merit further investigation, and (3) providing a clean sample without any peculiar or problematic spectra, so that further studies, based on the detailed stellar parameters and chemical abundances produced by Galah, can be more reliable. The method that we use to identify patterns or groups in the ``feature space'' to achieve the stated goals is unsupervised classification with t--SNE \citep{citeulike:3749741} reduction of spectral information. The main advantage of this approach is that we are more likely to detect various unfamiliar morphological features as well as the many known and expected peculiar stars. For a very nice overview of the vast range of classification and data mining techniques we refer the reader to \cite{2009ApJ...703.1061S}. 
 
The paper is organised as follows: the data reduction and overview of Galah spectra is described in Section \ref{data}, the classification procedure with a description of the employed techniques is detailed in Section \ref{class}, and the discovered classes of spectra are examined in Section \ref{classes} and \ref{young}. In Section \ref{catalogue} we present the structure of the catalogue with final classification results including supplementary information, and in Section \ref{visu} we briefly describe a web-based visualisation tool nicknamed \textit{Galah Explorer}, which displays the t--SNE projection map featuring various useful functionalities. We conclude with discussion in Section \ref{disc}.

\section{Data and reduction overview} \label{data}

	\subsection{Galah spectra}
	The Galah survey was the main driver for the construction of Hermes (High Efficiency and Resolution Multi--Element Spectrograph), a fibre--fed multi-object spectrograph on the 3.9 m Anglo-Australian Telescope (AAT). Its spectral resolving power (R) is about 28,000, and there is also an R = 45,000 mode using a slit mask. The spectrograph is fed via 400 fibres distributed over $\pi$ square degrees of sky. Taking into account the Galah magnitude limitation ($V = 14$), up to 392 stars can be observed simultaneously in that relatively small angle up to a Galactic latitude of $|b| \sim 28^{\circ}$. Hermes has four simultaneous non--contiguous spectral arms centred at 4800, 5761, 6610 and 7740 {\AA} (hereafter blue, green, red, and IR band), covering about 1000 {\AA} in total, including H$\alpha$ and H$\beta$ lines. The spectrograph is designed to have $\sim$ 10\% efficiency and to achieve SNR $\sim$ 100 per resolution element at $V = 14$ in a 1 hour exposure, resulting in measured RV errors $<$ 1 kms$^{-1}$ \citep{2015JATIS...1c5002S}.

	\subsection{Reduction pipeline}	\label{reduc}
	All the spectra subject to our analysis are reduced by the pipeline used in the Galah survey to produce fully calibrated spectra for subsequent stellar atmospheric parameter estimation \citep{2016MNRAS.tmp.1183K}. The reduction pipeline is based on reliable Iraf routines and other readily available software. After the Iraf--based reduction, a code that provides first estimates of radial velocity and three basic atmospheric parameters is run and normalization of the entire observed spectrum is done for each star (see Section 6 in \citealp{2016MNRAS.tmp.1183K}). For some spectra, processing by this code can fail due to various data issues, and such cases are excluded from further consideration. Otherwise, the values of the three parameters (T$_{\mathrm{eff}}$, log $g$, [Fe/H]), to which we refer in the text and which are colour coded in several figures, are produced by this code. These values are preliminary, but they are the best in terms of completeness for the currently analysed dataset. Improved stellar parameters determined by the Galah spectroscopic analysis pipeline are also available, using a combination of the spectral synthesis program Spectroscopy Made Easy (SME) \citep{1996AAS..118..595V,2016arXiv160606073P} and the data-driven approach The Cannon \citep{2015ApJ...808...16N}. The procedure is
summarised in \cite{2016arXiv160902822M} and will be detailed in \cite{asplund}. We are in the process of analysing these improved results.

	About 300,000 spectra have been taken to date, including various calibration exposures. However, we concentrate on $\sim$210,000 spectra recorded before 30th January 2016 and reduced with the Iraf reduction pipeline version 5.1. In the future, the same study will be extended to include additional spectra once they become available. For more details, we refer the reader to the thorough description of the reduction process \citep{2016MNRAS.tmp.1183K}.

\section{Classification} \label{class}
	
	We devise a custom classification procedure which is based on two independently developed methods, the novel dimensionality reduction technique t--SNE \citep{citeulike:3749741} and the renowned clustering algorithm DBSCAN \citep{citeulike:3509601}. Both are used more than once in an iterative approach to enable the most efficient classification and overview of our dataset. It should be noted that these purely mathematical methods are used extensively in various domains of research for unsupervised classification or clustering, and were not primarily intended for astrophysical purposes. The t--SNE/DBSCAN \textit{dimensionality reduction/clustering} combination was chosen out of the large variety of data-mining techniques because it is able to:
	\begin{itemize}
	\item handle inherently non-linear data (physics of line
formation in stellar photospheres) as opposed to traditional linear dimensionality reduction methods, e.g. PCA, MDS (Multi Dimensional Scaling; \citealt{young2013multidimensional}), LDA (Linear Discriminant Analysis; \citealt{Izenman2008}), CCA (Canonical Correlation
Analysis; \citealt{hotelling1936relations}), MAF (Min/max Autocorrelation Factors; \citealt{switzer1984min})
	\item  reveal the local as well as the global structure of the high--dimensional data in a single map
	\item alleviate the ``crowding problem'' that hampers many other non-linear techniques
	\item detect clusters in the projection map without a priori knowledge of their number and the form of their distribution function, contrary to some classical clustering methods, e.g. K-Means (\citealt{hartigan1979algorithm}), Gaussian Mixture Models (\citealt{marin2005bayesian})
	\end{itemize} 
		
	These advantages, supported by previous experience and a detailed performance comparison of t--SNE to other techniques by \cite{citeulike:3749741}, guided us towards the choice of algorithms for classification. However, we do not claim that these are the optimal methods, and an evaluation of the effectiveness of t--SNE/DBSCAN versus other methods in the context of the Galah dataset is beyond the scope of this study. Nevertheless, we are encouraged by our results in that the t--SNE/DBSCAN combination proved to be very useful and efficient in visualising and distinguishing different morphological groups of Galah spectra. 
	
	We first present both techniques in more detail and then focus on our custom classification procedure.

	\begin{figure*}[!htp]
  		\centering
  		\includegraphics[trim = 40mm 0mm 40mm 10mm, clip, width=1\textwidth]{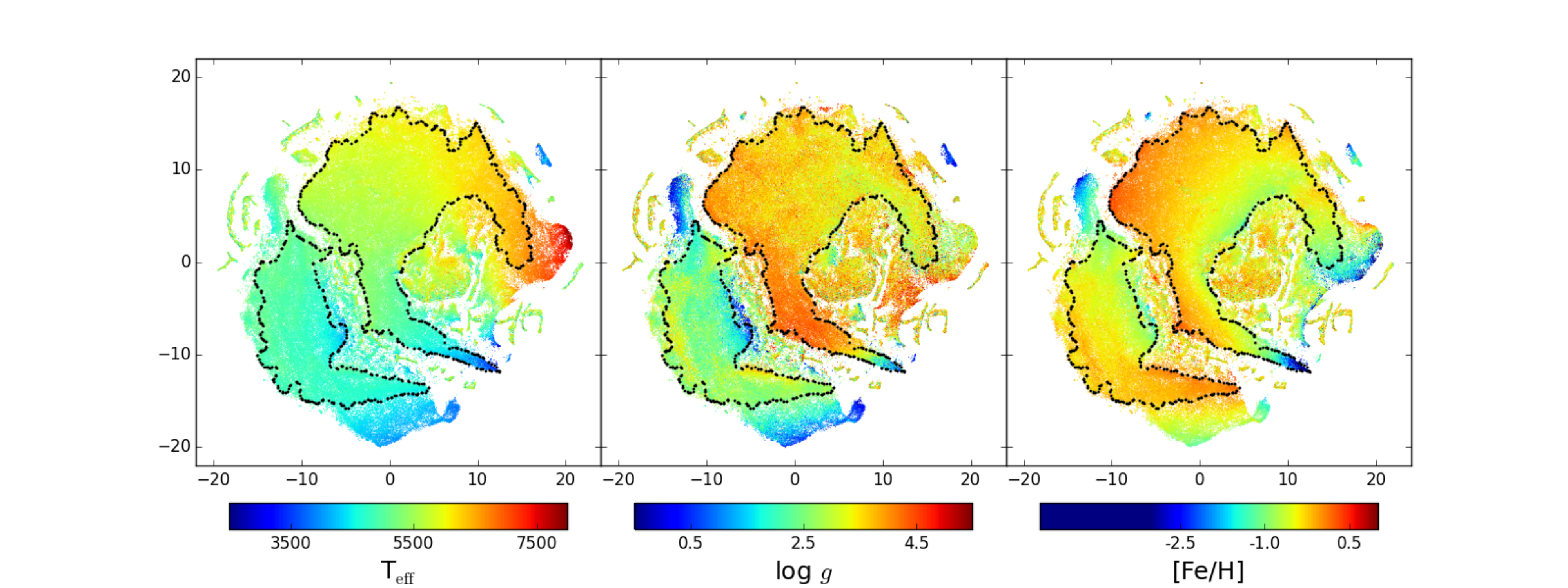}  		
		\caption{The first t--SNE projection of the whole working set containing 209,533 datapoints (spectra). The three panels feature T$_{\mathrm{eff}}$, log $g$, and [Fe/H] values for spectra (represented by the colour scale), measured by the Galah reduction pipeline (see Section \ref{reduc}). The lowest values of [Fe/H] are few and probably erroneous. The two areas encircled by black points are the two largest collections of the most appropriate DBSCAN mode ($Eps = 0.2$, $MinPts = 30$) for large scale cluster detection, which were removed in the third step of our classification procedure. The outer borders of the two collections are dotted, indicating the approximate shape which is not smooth and can be patchy on the inside as well. A relatively smooth distribution of parameters is clearly seen in the enclosed areas. Altogether DBSCAN defines 235 collections in this mode, which are not marked here. The axes of the panels have no physical meaning, they merely span the low--dimensional projection space.}
		\label{step12}
	\end{figure*}

	\subsection{t--SNE reduction of spectral information}
	The widely varying dimensionality and huge amounts of data in different domains of research necessitate some form of reduction and visualisation tools in order to efficiently extract information. t--SNE (t--distributed Stochastic Neighbour Embedding) can visualise any high--dimensional dataset by projecting each datapoint into a low--dimensional map, which reveals local as well as global structure of the data at many different scales. This is particularly suitable for high--dimensional data that lie on several different, but related, low--dimensional manifolds. An example in astronomy is a dataset of single-lined binary spectra of multiple spectral types shifted by different radial velocities. 
	
	The technique significantly improves the overall performance of its predecessor Stochastic Neighbor Embedding \citep{hinton2002sne}, mainly due to easier optimisation of the algorithm and a reduction of the tendency to crowd points together in the center of the projection map. For more information on the  performance of t--SNE on a wide variety of data sets and comparison with many other non-parametric visualization techniques, including Sammon mapping, Isomap, and Locally Linear Embedding, we refer the reader to the original paper \citep{citeulike:3749741}. We note that t--SNE has already been used quite recently in the astronomical community \citep{2016IAUS..317..336M,2016ApJS..225...31L,2016arXiv160903826V}. A brief introduction to the technique including formulae and text adapted from \cite{bib:pezzotti:2016} is given in the following paragraphs.
	
	A low--dimensional representation of high--dimensional data is achieved by optimally positioning datapoints in the projection map. For this purpose, t--SNE defines the similarity between $N$ datapoints in the original high--dimensional space $X$ and in the projection space $Y$, described by symmetric joint--probability distributions $P$ and $Q$, respectively. More precisely, the pairwise similarity between datapoints is modelled by the probability that one datapoint would pick another one as its neighbour, which depends on the probability density under a Gaussian in space $X$, whereas a \textit{Student's t-Distribution} is used in space $Y$:

	\begin{align}
		p_{ij} &= \frac{p_{j|i}+p_{i|j}}{2N} \\
		\mathrm{where }	\qquad  p_{j|i} &= \frac{\exp(-\Vert \textbf{x}_i - \textbf{x}_j \Vert^2 / 2 \sigma_i^2)}{\sum_{k \neq i} \exp(-\Vert \textbf{x}_i - \textbf{x}_k 		\Vert^2 / 2 \sigma_i^2)}
	\end{align}
	
	for space $X$ and
	 
	\begin{align}
		q_{ij} &= ((1+ \Vert \textbf{y}_i - \textbf{y}_j \Vert^2)Z)^{-1} \\
		\mathrm{where } \qquad	Z &= \sum_{k = 1}^N \sum_{l \neq k}^N (1 + \Vert \textbf{y}_k - \textbf{y}_l \Vert^2)^{-1}
	\end{align}
	
	for space $Y$. 
	
	The $\sigma_i$ is computed for each datapoint $\textbf{x}_i$ so that the effective number of its neighbours corresponds to the fixed user-defined parameter $\mu$ (perplexity):
		
	\begin{equation}
	\mu = 2^{- \sum_j^N  p_{j|i} \log_2 p_{j|i}}	
	\end{equation}
	
	In regions of space $X$ with a higher data density, $\sigma_i$ tends to be smaller than in regions of lower density. The importance of modelling the separations between $\textbf{x}_i$ and $\textbf{x}_j$ does not depend on their absolute distance in space $X$ as long as they are close to each other relative to $\sigma_i$. Likewise, the size of the local neighbourhood of $\textbf{x}_i$ depends strongly on $\sigma_i$, and the similarity measure $p_{j|i}$ becomes almost infinitesimal for $\textbf{x}_j$ at the distance of several $\sigma_i$ due to the nature of the Gaussian distribution. These properties effectively define a soft border between the local and global structure of the data.
	
	The novel element in computing the joint--probability distribution $Q$ in space $Y$ is in using a normalised \textit{Student's t-Distribution} kernel with a single degree of freedom. The heavy tails of this distribution put more space between the moderately dissimilar points than a Gaussian would. As a result, there is more projection space available so that $\textbf{y}_i$ and $\textbf{y}_j$ can model the local structure of the corresponding $\textbf{x}_i$ and $\textbf{x}_j$ more accurately. 	
	
	Having defined the joint--probability distributions $P$ and $Q$, t--SNE aims to optimally position the points in space $Y$ by minimising the non--convex cost function $C$ given by a simple measure of (Kullback--Leibler) divergence between probability distributions: 
	
	\begin{align} \label{kul}
		C(P, Q) = KL(P \Vert Q) = \sum_i \sum_j p_{ij} \log \frac{p_{ij}}{q_{ij}}
	\end{align}
	
	This is achieved with an iterative gradient descent using a stochastic element to avoid local minima and governed by the gradient of the above  divergence:
	
	\begin{align}
		\frac{\partial C}{\partial \textbf{y}_i} &= 4 \sum_{i=1}^N (F_i^{\mathrm{attr}} - F_i^{\mathrm{rep}}) \\
		&= 4 \sum_{i=1}^N ( \sum_{j \neq i}^N p_{ij} q_{ij} Z (\textbf{y}_i - \textbf{y}_j) - \sum_{j \neq i}^N q_{ij}^2 Z (\textbf{y}_i - \textbf{y}_j) )
	\end{align}

	The gradient descent can be seen as an N--body simulation, where each datapoint exerts an attractive and a repulsive force on all the other points ($F_i^{\mathrm{attr}}, F_i^{\mathrm{rep}}$). In this respect, t--SNE gradient has the advantage of strongly repelling datapoints modelled by small pairwise distances in space $Y$ which are otherwise very dissimilar in space $X$, but these repulsions do not increase to infinity as in some other attempts to address the crowding problem (see Sections 3.2, 3.3 and Figure 1 from \citealp{citeulike:3749741}). 
	
In this work, we are dealing with over 200,000 spectra which can be viewed as datapoints in space $X$ of dimensionality over 13,000. The high computational complexity introduced by employing t--SNE on our growing dataset requires that we make use of the Barnes--Hut t--SNE \citep{2013arXiv1301.3342V}, an evolution of the t--SNE algorithm that introduces different approximations to reduce the computational cost from $O(N^2)$ to $O(N \log(N))$ and the memory complexity from $O(N^2)$ to $O(N)$. 

When computing the t--SNE embedding, we always: (a) project our dataset into two--dimensional space, (b) set the Barnes--Hut parameter $\theta$ to 0.5, and (c) unless otherwise specified, set the perplexity to 30, a value that has generally proven to be most effective for our purpose. Smaller values of perplexity produce sparser projection maps with denser collections of points and larger values produce more evenly covered projection space but with less pronounced separations between distinct groups. The choice of two--dimensional space compared to three dimensional is pragmatic, since we only use one image (projection map) for visual inspection in the 2D case, whereas in 3D we would need to inspect three projected planes or use some advanced 3D visualisation tool. We plan to explore the 3D option in the future, as it potentially preserves more information of the underlying structure of data. 

The t--SNE procedure can be summarized in: (1) converting Euclidean distances in space $X$ to pairwise similarities (often computationally the most intensive part), (2) sampling map points randomly from an isotropic Gaussian with small variance that is centered around the origin of projection space $Y$, and (3) initialising the gradient descent with a fixed number of iterations (usually $1000$).

\subsection{DBSCAN clustering}

Density-based spatial clustering of applications with noise (DBSCAN) is a data clustering algorithm relying on a density--based notion of clusters (collections) and designed to discover any arbitrary shape of collections of points in some space. It defines clusters from densely packed points -- those that have many nearby neighbours -- and rejects those points that lie alone in low-density regions (outliers). DBSCAN is one of the most common clustering algorithms and also one of the most cited in scientific literature.

There are two input parameters to DBSCAN method that have to be set by the user, $minPts$ and $\varepsilon$. Furthermore, the points are classified into three groups: core points, \mbox{(density--)reachable} points, and outliers, defined as follows:
\begin{itemize}
\item A point $u$ is a core point if at least $minPts$ points are within distance $\varepsilon$ of it (including $u$), and by definition these are directly reachable from $u$, whereas no points are directly reachable from a non-core point.
\item A point $v$ is reachable from $u$ if there is a path $u_1$, ..., $u_n$ with $u_1 = u$ and $u_n = v$, where each $u_{i+1}$ is directly reachable from $u_i$ (all the points on the path must be core points, with the possible exception of $v$, in which case $v$ is a border point).
\item All points not reachable from any other point are outliers.
\end{itemize}

To find a collection of points, DBSCAN starts with an arbitrary point $u$ and retrieves all points density-reachable from $u$ wrt. $\varepsilon$ and $MinPts$. If $u$ is a core point, this procedure yields a collection. If $u$ is a border point, no points are density-reachable from $u$ and DBSCAN visits the next point. 

In this study, DBSCAN is employed merely as a tool for automatic detection of distinct collections in the projection map produced by t--SNE, without any bias apart from the manual selection of $\varepsilon$ and $MinPts$. By definition of the t--SNE projection map, datapoints that are similar to each other should be closely packed together, thereby, in DBSCAN's terminology, forming a collection which can be detected and labelled.

\subsection{Classification procedure} \label{classp}

	In the process of finding the most objective and efficient way of classifying Galah spectra, we have established a procedure which is a combination of automatic and manual processing and inspection of our data. The current Galah reduction pipeline includes several stages (see Section \ref{data}) and we only retain those spectra that pass radial velocity determination and normalisation, which are in our analysis the two key properties for optimal use of t--SNE dimensionality reduction. In principle, we could use the whole dataset of spectra before any kind of reduction, but in that case the most important features driving our low--dimensional embedding (projection map) would be a result of missing wavelength calibration, flux normalisation, radial velocity determination, etc. In essence, we would be concerned with properties that are less scientifically compelling and can be reliably enough accounted for with standard reduction procedures. After this selection, the number of spectra that remain is 209,533, and we join together the four spectral bands to produce the so--called datapoints of our working set.
		
		Taking the whole practically usable range of red, green, blue, and IR band (excluding the strongest telluric features in the latter), the number of normalised flux values, which are basically pixels or original dimensionality of datapoints, amounts to 13,600 per datapoint (spectrum). At such high dimensionality multiplied by 209,533 datapoints, the computational cost despite the Barnes--Hut implementation of t--SNE can still turn out to be overwhelming or impractical (taking over ten days to compute on an Intel(R) Xeon(R) 2.60GHz CPU, using over 80GB of memory). To ease the whole process, in terms of memory consumption and computation time, and to produce a projection map most suited for the purpose of this work, we make use of the following scheme:

\begin{figure*}[!htp]
  		\centering
  		\includegraphics[trim = 40mm 0mm 40mm 10mm, clip, width=1\textwidth]{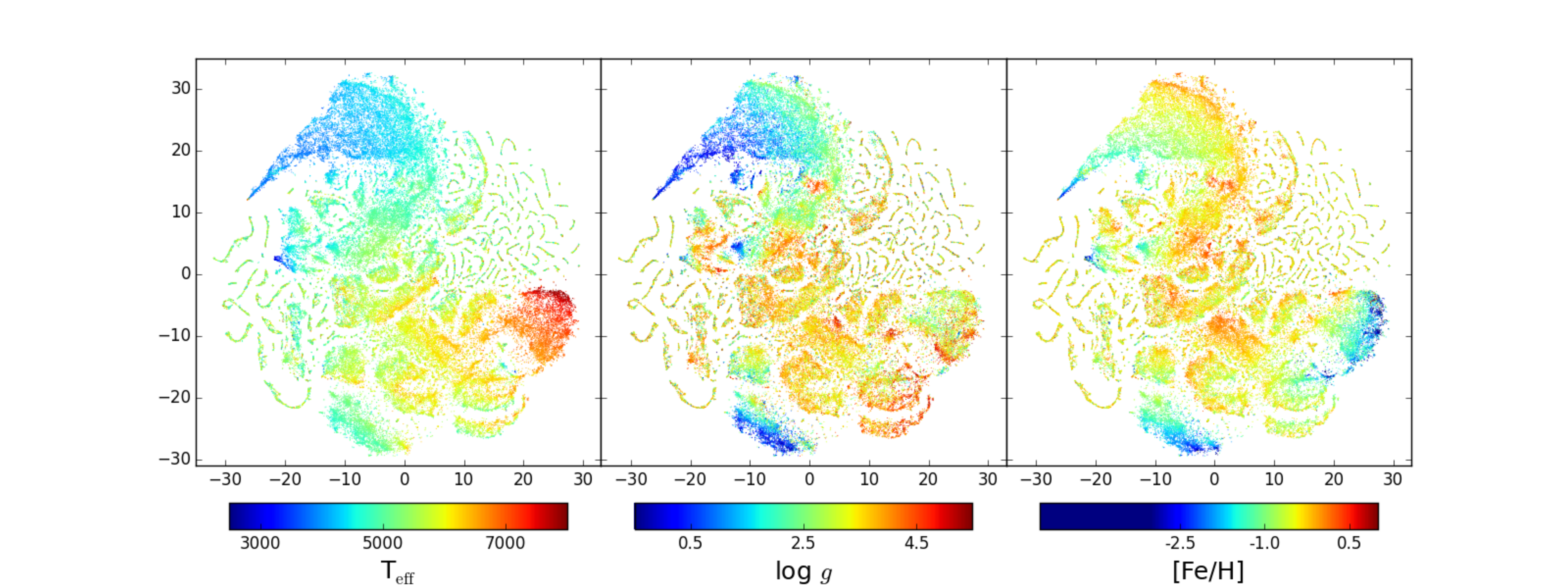} 
		\caption{Same as Figure \ref{step12} but for the second t--SNE projection map of the filtered working set containing 76,938 datapoints (spectra), i.e. those that are not part of the two large collections marked in Figure \ref{step12}.}
		\label{step3}
	\end{figure*}

	\begin{enumerate}
	
		\item \textbf{First t--SNE projection}: the first projection is computed for the whole working set (209,533 spectra). We use only the following wavelength ranges: $4850-4880$ {\AA}, $5750-5780$ {\AA}, $6550-6580$ {\AA}, and $7730-7760$ {\AA}. These ranges are selected so as to include the more diagnostic parts of each spectral band, with equal contribution from all of them, amounting to $2400$ as dimensionality of datapoints. The map resulting from this first projection can be used as is, for this is the most basic and objective clustering of all datapoints from our working set. However, a significant portion of spectral information is missing due to our cut in wavelength range. There is also a practical caveat, in that although the projection map is fairly homogeneously populated, it is also very dense, making the smaller scale clustering, the one we are most interested in, more difficult to recognise. The next steps alleviate these issues. 
		
		\item \textbf{DBSCAN large scale cluster detection}: DBSCAN input parameters are set in a way that a few large collections of datapoints are defined across the projection map. This is done in order to select those collections that presumably contain only the ``normal'' spectra. In our experience, well behaved spectra are usually clustered together in one or a few large areas of the map where the atmospheric parameters (T$_{\mathrm{eff}}$, log $g$, [Fe/H]) are continuously distributed. 
		
		\item \textbf{Select and filter out collections of ``normal'' spectra}: It is here that manual interaction is most important, since we are rejecting (to our analysis) less interesting  datapoints and even if being very careful, we can unwittingly discard some of the desired spectra from further consideration. The map of the first t--SNE projection with DBSCAN clustering on large scale is shown in Figure \ref{step12}. The two largest collections outlined by black dotted line, amounting to 137,155 datapoints, are rejected for containing presumably only ``normal stars'', and 76,938 remaining datapoints are considered in the next steps.		
				
		\item \textbf{Second t--SNE projection}: the second projection is computed for the subsample of the working set (76,938 spectra), resulting from the previous step. Due to the smaller number of datapoints compared to the first projection, it is now feasible to operate with all practically usable flux values (13,600) from the four spectral bands: $4730-4880$ {\AA}, $5670-5850$ {\AA}, $6500-6710$ {\AA}, and $7725-7865$ {\AA}. The projection map is shown in Figure \ref{step3}, and serves as the final basis for our selection and analysis of peculiar spectra. Some ``normal'' spectra are still present in this map, but the largest portion should belong to all the peculiar objects that we are interested in. Their small scale structure, which was hidden in the first t--SNE projection, is now reflected in the large scale structure, and also more easily discernible due to the overall fewer datapoints in the map, as evident from comparing Figures \ref{step12} and \ref{step3}.
		
		\item \textbf{DBSCAN small scale cluster detection}: DBSCAN input parameters are set in a way that the defined collections correspond to relatively small and dense regions in the map which represent distinct morphological classes of spectra. In our experience, there is no unique parameter set for DBSCAN that would allow us to properly select the various collections, so many sets of parameters are tried and the corresponding DBSCAN results (hereafter DBSCAN modes), producing different sizes, shapes, and numbers of collections, are available for inspection in the next step. 
		
		\item {\bf Select relevant/categorical collections and assign classification categories/flags}: The final step involves manual overview of individual spectra in different collections with the help of our visualisation tool presented in Section \ref{visu}. The goal is to find the best collection from different DBSCAN modes that fully encompasses the manually examined spectra belonging to a distinct category. Some outliers in terms of a chosen category will usually be present, so the final results should be regarded as a list of candidate members of a certain category. We do not assign a quantitative probability of membership as it is beyond the scope of this study. Furthermore, the selected collections from different DBSCAN modes might sometimes overlap, so one spectrum might be assigned to two categories, in which case it can be viewed as a candidate member of both.

	\end{enumerate}

\section{Morphological classes of spectra} \label{classes}

	Table \ref{simbad} lists 6 distinct categories that were defined using the classification procedure described in Section \ref{classp}. This classification is not limited strictly to peculiar objects having spectra without a counterpart in the library of synthetic spectra, although they remain the principal motivation for this work. Rather it is a search for any coherent group in the projection map, from which anyone can pinpoint their category of interest. They range from larger collections of points (spectra), like the \textit{Hot stars} category, to smaller ones that mostly contain problematic spectra, having usually one, albeit very prominent, feature (e.g. a strong emission spike). 
	
\begin{figure*}[!htp]
  		\centering  		
  		\includegraphics[trim = 45mm 25mm 45mm 35mm, clip, width=0.8\textwidth]{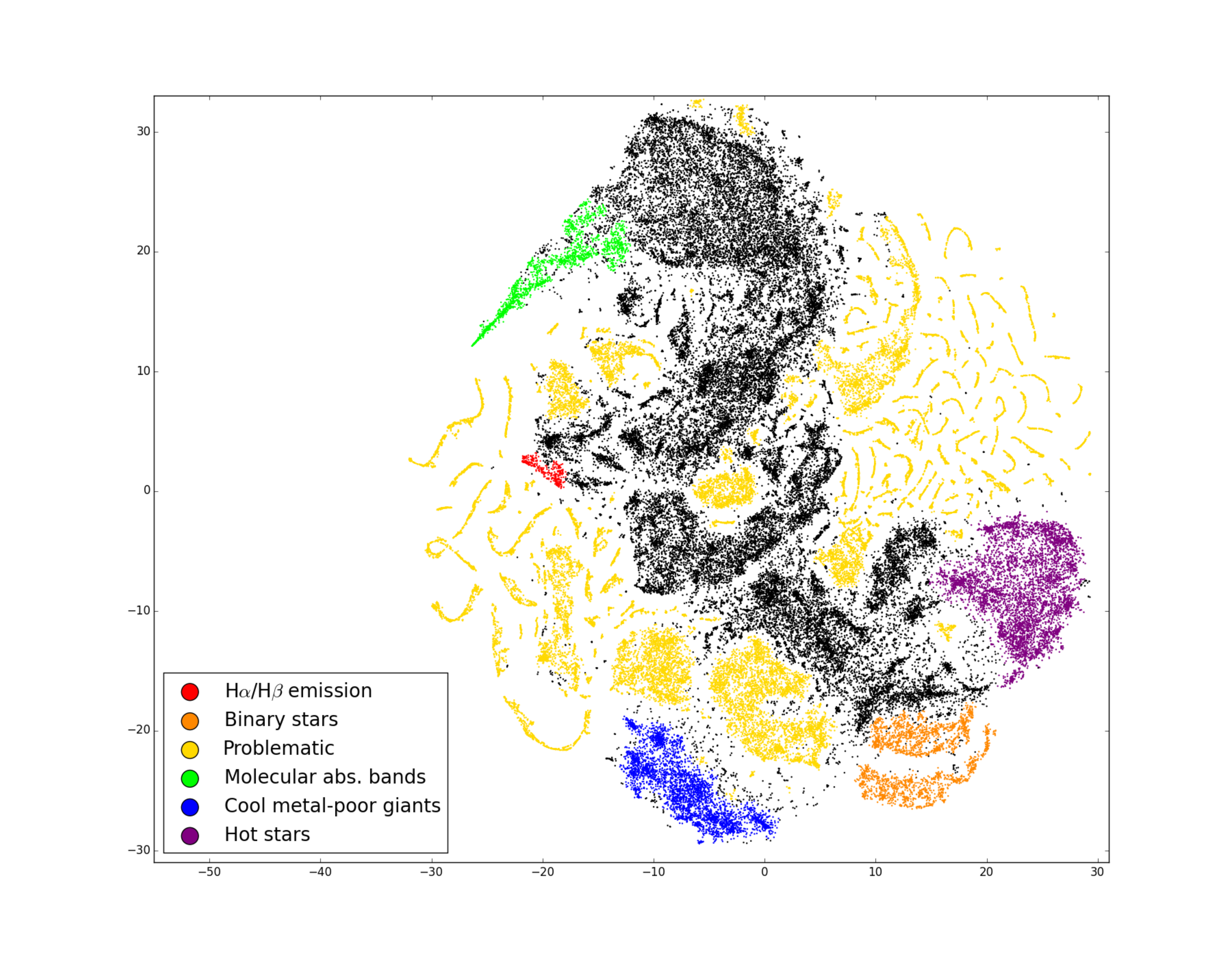}
		\caption{The result of the classification procedure, based on the projection map in Figure \ref{step3}. Collections of spectra assigned to distinct categories are flagged (colour coded), the rest are black. The axes of the panels have no physical meaning, they merely span the low--dimensional projection space.}
		\label{step4}
	\end{figure*}

	The projection map that was used to search for and define the 6 general categories is presented in Figure \ref{step3}. The overall distribution of parameters T$_{\mathrm{eff}}$, log $g$, and [Fe/H] in the three panels indicates their importance in feature space. T$_{\mathrm{eff}}$ is clearly the principal discriminant with a gradient over the whole projection map, followed by [Fe/H] and log $g$ which influence the distribution of points inside larger and well separated collections. The collections that represent distinct categories are marked in Figure \ref{step4}. Some of them are characterised by a certain strong spectral feature, hence the three main stellar parameters can be well mixed within the collection, in addition to being erroneous in cases where such features prevent a reliable estimate of their values. 
	
	For all targets corresponding to spectra with an assigned classification category, a search by coordinates inside 1 arcsec radius was performed on the SIMBAD database. The most common \textit{Main type} and \textit{Other type} property of the matched SIMBAD objects are listed in Table \ref{simbad}. In the following paragraphs, categories are described individually, with several issues related to observations and reduction combined in the category \textit{Problematic}.
	
	\subsection{Hot stars}
\begin{figure*}[!htp]
  		\centering
  		\includegraphics[trim = 20mm 20mm 20mm 30mm, clip, width=0.92\textwidth]{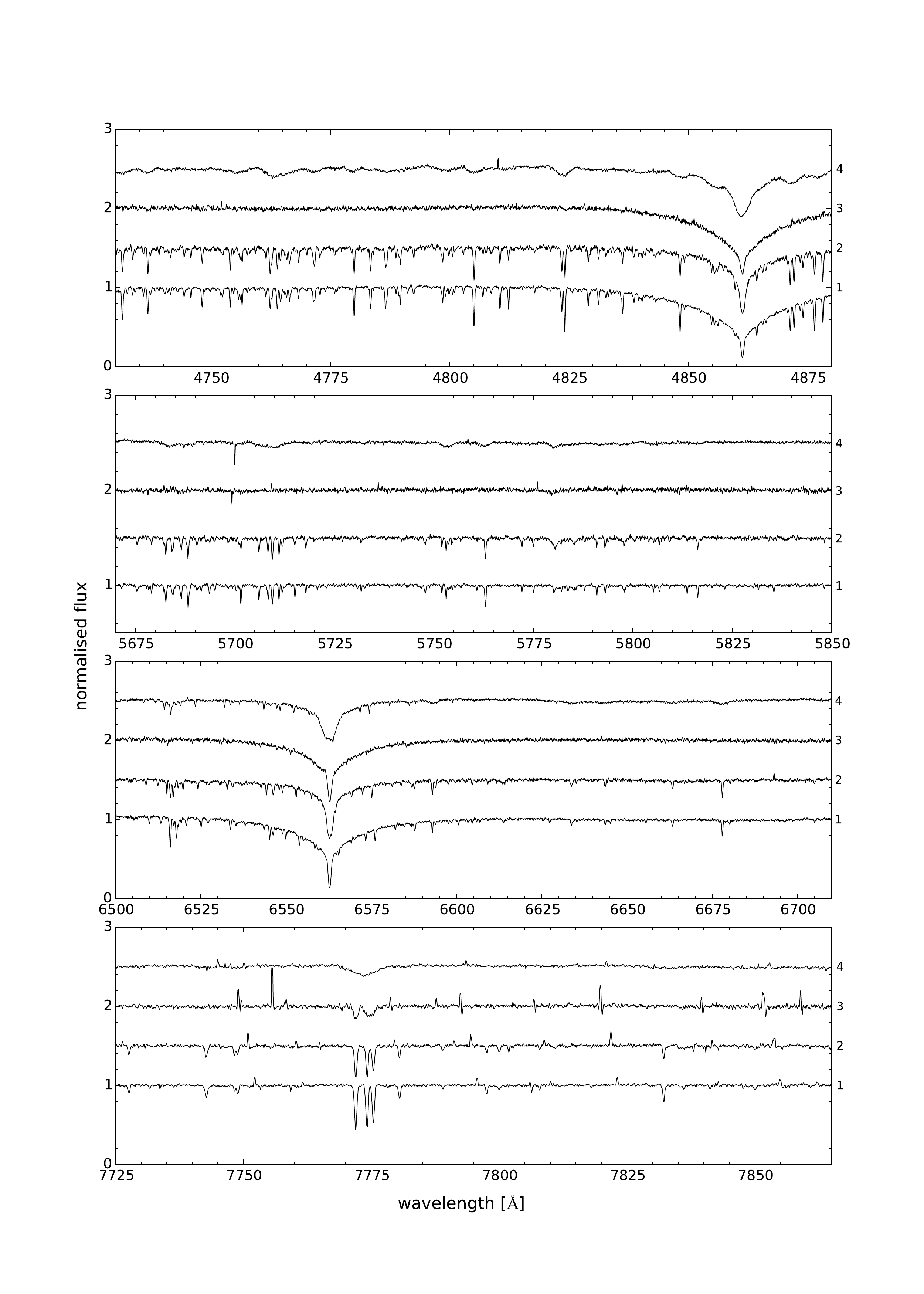}  		
		\caption{Examples of spectra of \textit{Hot stars} category. Separate panels represent spectra from different Hermes spectral bands but for the same stars. Vertical spacing between spectra in each panel is adjusted for clarity. From bottom to top in each panel, spectra are labelled in sequence on the right axis, corresponding to spectrum number in Table \ref{specs}.}
		\label{hot}
	\end{figure*}

	A large collection at the right part of the map in Figure \ref{step3} contains mostly early type stars, with temperatures well above solar, characterised predominantly by widened wings of H$\alpha$ and H$\beta$ absorption lines. We observe a smooth transition of temperatures inside this collection, ranging from about $6500$ up to $8000$ K (upper limit of the grid of synthetic templates, see \citealp{2016MNRAS.tmp.1183K}). The distribution of metallicity is also very smooth, along an axis perpendicular to temperature, while the surface gravity is more patchy, with dwarfs more clustered in some parts and giants more dispersed throughout the collection. Examples of spectra in this category with different metallicity and temperatures are shown in Figure \ref{hot}.
	
	\subsection{Cool metal-poor giants}
\begin{figure*}[!htp]
  		\centering
  		\includegraphics[trim = 20mm 20mm 20mm 30mm, clip, width=0.92\textwidth]{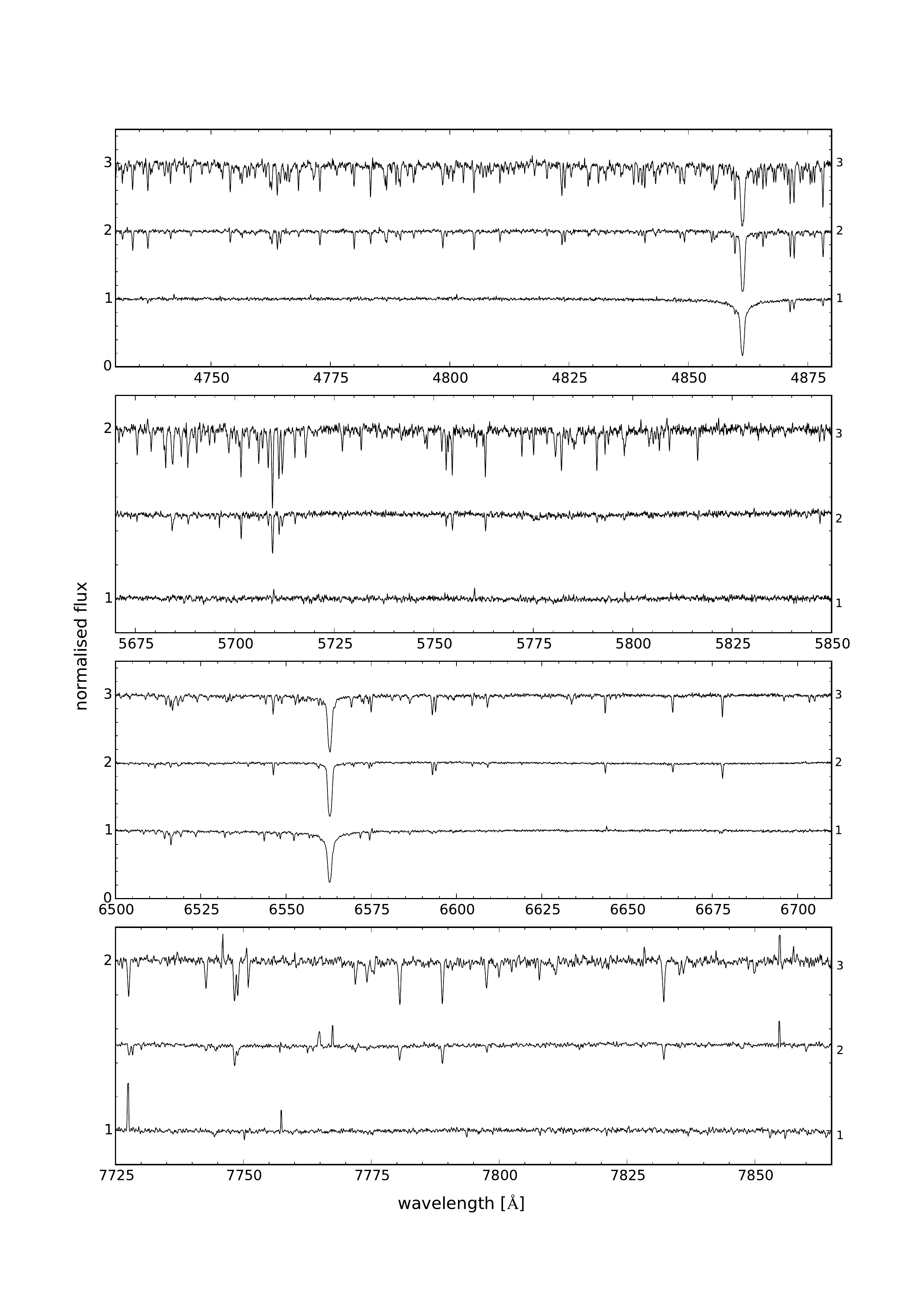}  		
		\caption{Same as Figure \ref{hot} but for the \textit{Cool metal-poor giants} category.}
		\label{cool}
	\end{figure*}

	The collection in the bottom part of the projection map features mostly late type stars with a measured metallicity well below solar value ($-4.5 <$ [Fe/H] $< -0.5$, see Figure \ref{cool}). The distribution of surface gravity is clearly seen, with the majority of stars being giants, and most with temperatures in the range from $4000$ K to a little above solar. The available records from SIMBAD support these claims, with 162 stars classified as \textit{Red Giant Branch star}, 26 as \textit{Possible Red Giant Branch star}, and 10 as \textit{Variable Star of RR Lyr type}.
	
	\subsection{Stars with molecular absorption bands}
\begin{figure*}[!htp]
  		\centering
  		\includegraphics[trim = 20mm 20mm 20mm 30mm, clip, width=0.92\textwidth]{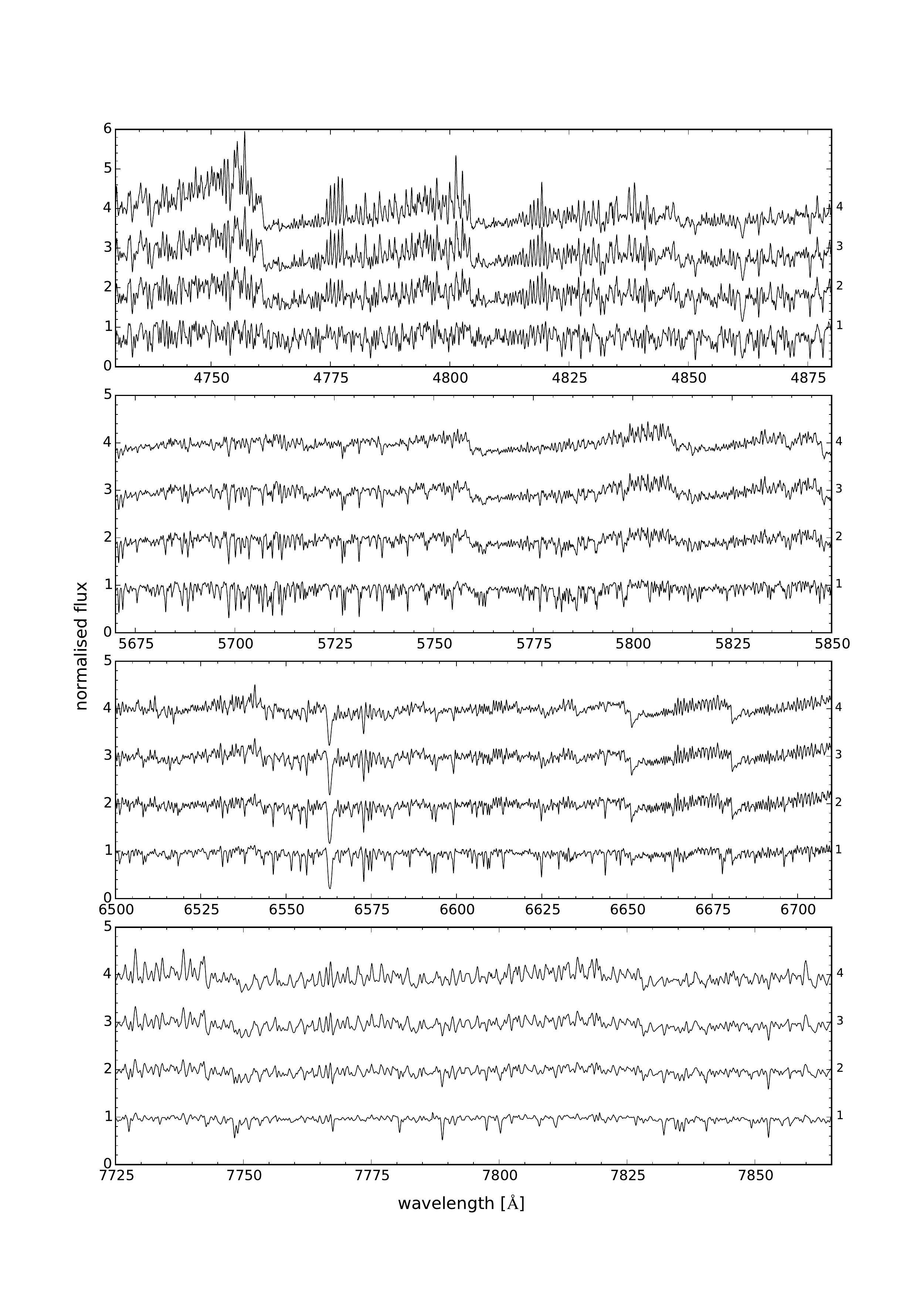} 
		\caption{Same as Figure \ref{hot} but for the \textit{Molecular absorption bands} category.}
		\label{molecular}
	\end{figure*}

	The upper left part of the map in Figure \ref{step3} contains a region populated by spectra with strong molecular absorption bands. It is well isolated on one end but still connected to the rest of the late type stars on the other end. The temperatures in this collection are mostly low as expected, but not unquestionable, as is nicely demonstrated by the very tip of this area on the left side, where we find the strongest absorption bands, while the temperatures derived for some of these spectra are much too high (above $6500$ K). The increase in strength of absorption bands nicely follows the direction from this extreme end to the larger region of late type spectra (from top to bottom on panels in Figure \ref{molecular}). The surface gravities and metallicities in this collection may also be problematic, due to the known problem of producing reliable synthetic templates for such stars. 
	
	\subsection{Binary stars}
\begin{figure*}[!htp]
  		\centering
  		\includegraphics[trim = 20mm 20mm 20mm 30mm, clip, width=0.92\textwidth]{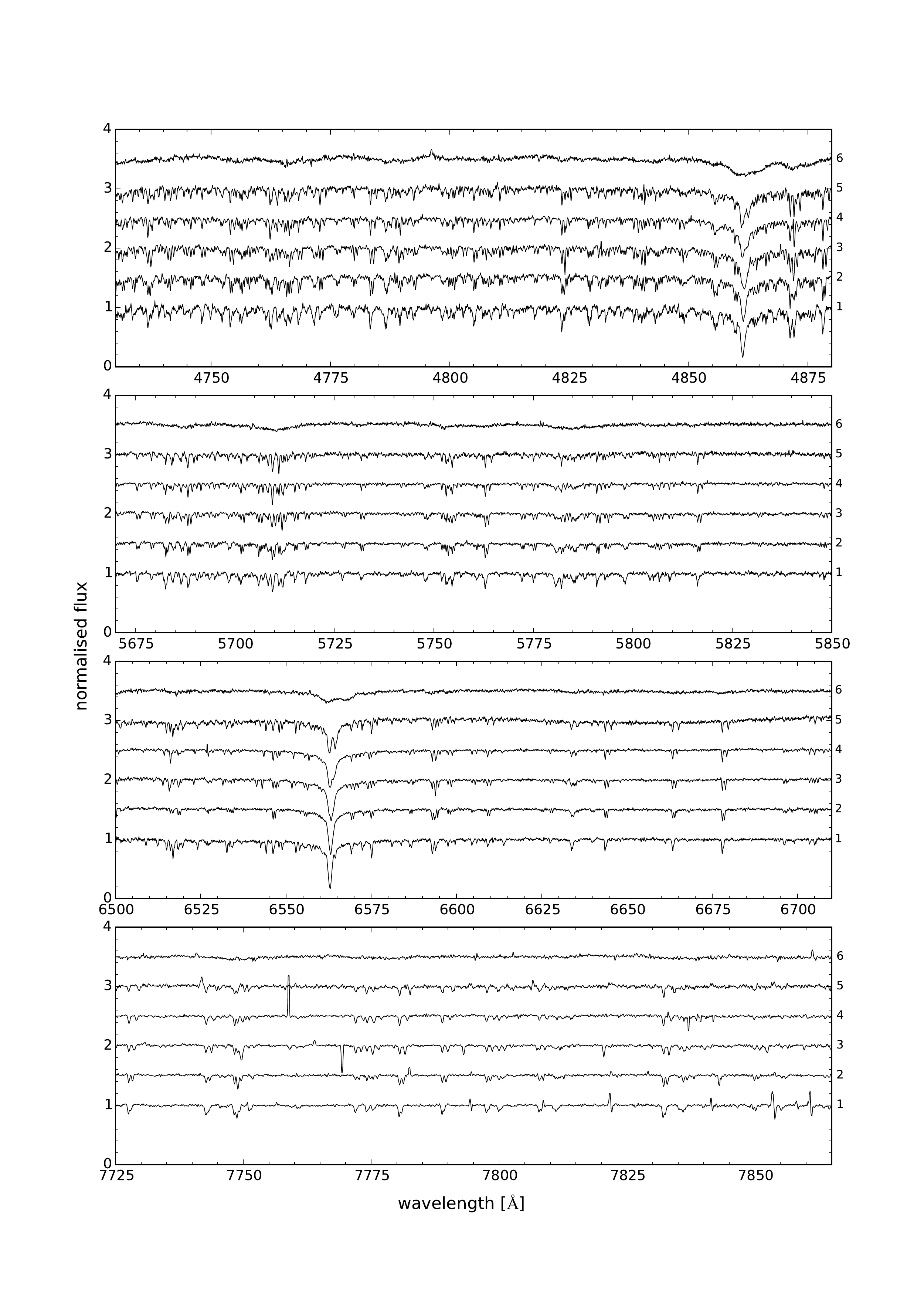}
		\caption{Same as Figure \ref{hot} but for the \textit{Binary stars} category.}
		\label{bin}
	\end{figure*}

	Multiple stars, of which the majority represents double-lined spectroscopic binaries (SB2), are found in the bottom right part of the projection map, clustered in two well separated collections. The main and most obvious difference between them is the position of the stronger of the two components in terms of the equivalent width of absorption lines. For the lower group the stronger component is positioned blueward, while for the upper group it is redward. Although the distinction is not physically significant, it is evidently morphologically important. Following the arc-like shape of the two collections from left to right, the spectra show a progressively larger radial velocity separation of the two components, from almost blended double lines to those separated by as much as $150$ kms$^{-1}$. Examples of spectra of stars in this category are shown in Figure \ref{bin}. The top star is a W Uma star and the spectra are shown with increasing radial velocity separation of the two components from bottom to top. We have some indications of binarity in Table \ref{simbad}, although only for a handful of stars, the other candidates being currently unknown for their binary nature according to the SIMBAD database. The same search by coordinates as performed on SIMBAD reveals no systems in \cite{2004AA...424..727P} and 2 systems in \cite{2001AJ....122.3466M} catalogues. By visual inspection, some SB3 candidates and W UMa type SB2s are also a part of this collection, although they are not isolated enough in the projection map to be labelled separately.
	
	The number of spectra in this collection represents around 1\% of the investigated Galah dataset. However, the true number of such objects is doubtlessly larger, as there are many factors hindering their detection: SB2 with blended lines, exclusion of potential candidates in the third step of our classification procedure or classification as \textit{Problematic} due to a stronger spectral feature. The simulation from \cite{2010AJ....140..184M} performed for SB2 analysis of RAVE spectra found that the detection rate should be fairly high ($\sim 80$\%) for systems with orbital periods shorter than $\approx 100$ days. The limiting line separation $\triangle v_{orb} \approx 50$  kms$^{-1}$ for RAVE (near-IR, SNR $\sim$ 45, $R \sim 7500$) should be less for Galah due to the higher resolution and signal-to-noise ratio of spectra and so the detection of longer period systems should be greatly improved. Indeed, the smallest separations among the detected binaries in this collection are $\triangle v_{orb} \approx 15$ kms$^{-1}$.
	
	\subsection{H$\alpha$/H$\beta$ emission}
\begin{figure*}[!htp]
  		\centering
  		\includegraphics[trim = 20mm 20mm 20mm 30mm, clip, width=0.92\textwidth]{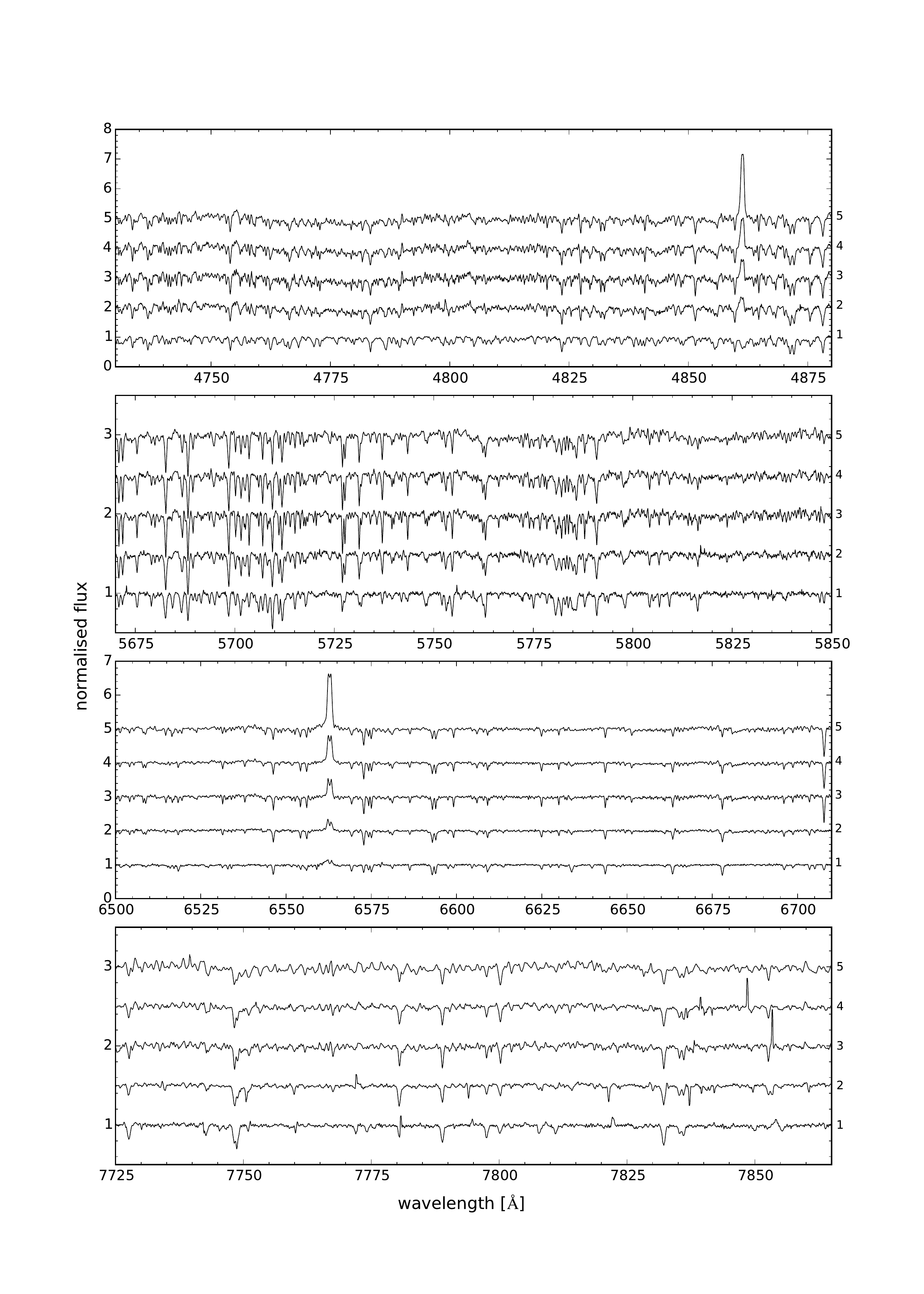}  		
		\caption{Same as Figure \ref{hot} but for the  \textit{H$\alpha$/H$\beta$ emission} category.}
		\label{hahb}
	\end{figure*}

	Emission-type stars often feature diverse profiles in H$\alpha$ and H$\beta$ emission lines, indicative of young stars, cataclysmic variables, symbiotic stars, stars with massive outflows or inflows, and many other types of active objects. The shapes of emission profiles can be described by meaningful morphological, and possibly also physical, categories as demonstrated by \cite{2015AA...581A..52T}. In this collection, the diverse profiles (double peaks, emission superimposed on absorption, P-Cygni, and others) are presented together, as they are relatively few and also not clearly separated in the projection map. The H$\alpha$ emission line is mostly present and often accompanied by a similar profile shape of H$\beta$ line. In some cases, molecular absorption bands and the lithium absorption line are clearly visible, all together are indicative of cooler, younger, and active stars. Examples of spectra in this category are in Figure \ref{hahb}.
	
	\subsection{Problematic}
\begin{figure*}[!htp]
  		\centering
  		\includegraphics[trim = 20mm 20mm 20mm 30mm, clip, width=0.92\textwidth]{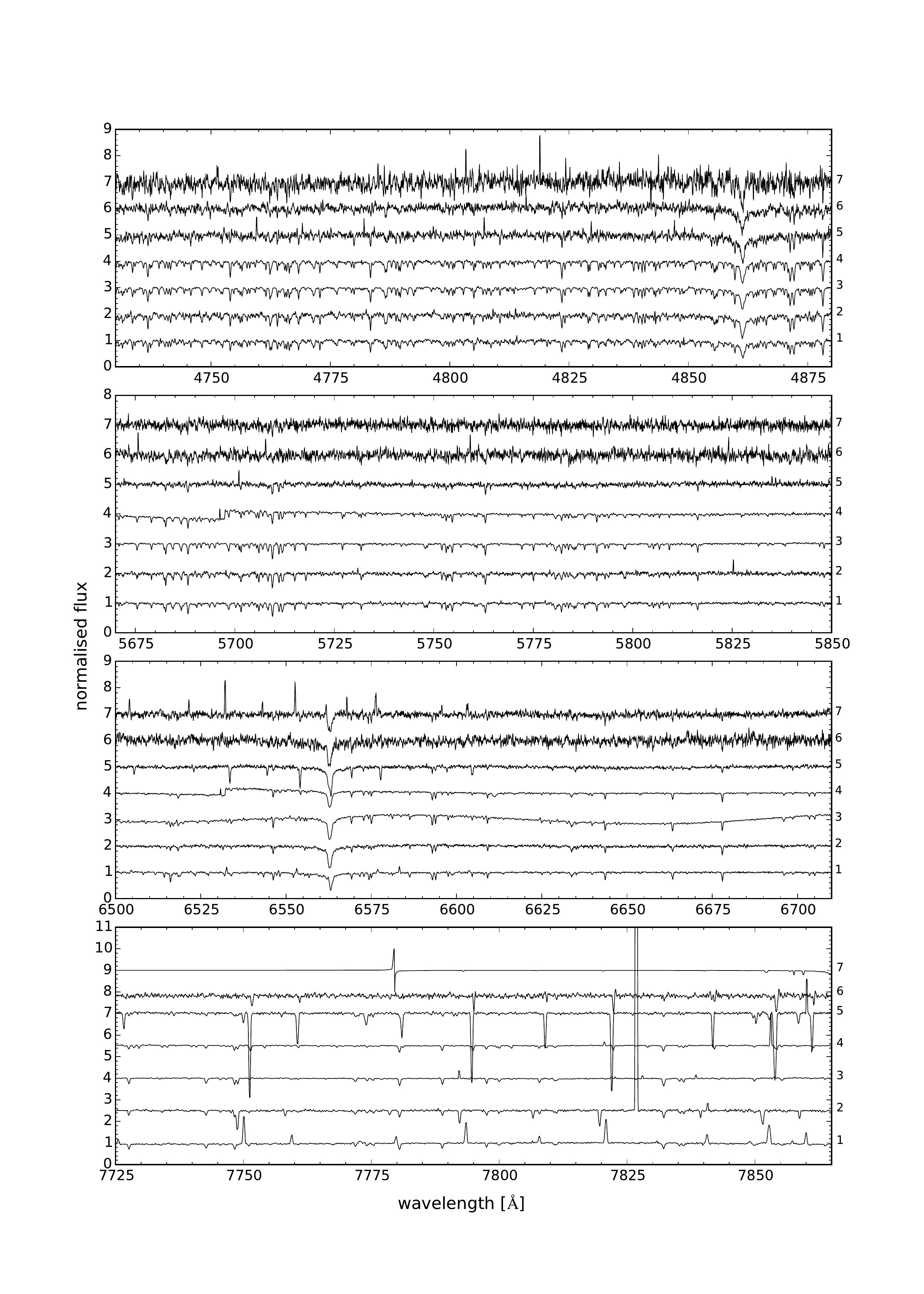} 		
		\caption{Same as Figure \ref{hot} but for the \textit{Problematic} category. Subcategories of spectra in the panels from bottom to top: emission spikes, strong emission spike, oscillating continuum, oscillating continuum, negative flux, low continuum, and low continuum.}
		\label{problem}
	\end{figure*}

	This collection is very diverse as it assembles together spectra which are in some peculiar, and generally undesirable fashion, affected in either observations or data reduction. \textit{Emission spikes} (5352 spectra) are most often present in the IR band and sometimes, but less strongly, in the red band, and are probably due to under-subtracted sky lines. The left and right parts of the map, shaped by low density snake--like collections, represent spectra with one \textit{strong emission spike} (9599) in the IR band. A normalisation issue in the form of an \textit{oscillating continuum} (2025) in the red band is also very common. \textit{Negative flux} (2078) is most often present in the IR band, followed by the blue band and less often in the red band, and might be due to sky over-subtraction. There is one more quite interesting, but less frequent, reduction effect in the IR band. This is in the form of very \textit{low continuum} (41) which is either at $\sim 0.3$ or close to and below zero level, often accompanied by strong oscillating features. These subcategories follow each other in Figure \ref{problem} from bottom to top in each panel. Most spectra in this category are well behaved in all aspects apart from the described issues, and their automatic detection without manual inspection is very helpful for the iterative development and improvement of our reduction pipeline \citep{2016MNRAS.tmp.1183K}.

\def\arraystretch{1.6}
\begin{table*}
\caption{Classification categories based on the general projection map (see Section \ref{classes}). The columns give the classification category, number of classified spectra, and the most common SIMBAD \textit{main types} and \textit{other types}. SIMBAD defines a \textit{main type} for each astronomical object in its database, and usually several \textit{other types} generally inferred from its identifiers. For the last two columns, only the five most common types are listed, excluding the less interesting type \textit{Star}. The full table is available in the electronic version. \label{simbad}}
\footnotesize
\centering
\begin{tabularx}{\linewidth}{ lrXX }
\hline\hline
Category & N & Main type & Other types\\
\hline
{\bf Hot stars} & 4130 & Star in Cluster (20), Variable Star of RR Lyr type (13), Variable Star (12), Variable Star of delta Sct type (12), Eclipsing binary of Algol type (detached) (8) & Infra-Red source (1486), Variable Star (41), Star in Cluster (26), Rotationally variable Star (19), Variable Star of RR Lyr type (13)\\
{\bf Cool metal-poor giants} & 2784 & Star in Cluster (371), Red Giant Branch star (162), Possible Red Giant Branch star (26), Variable Star of RR Lyr type (10), High proper-motion Star (8) & Infra-Red source (695), Star in Cluster (552), Red Giant Branch star (165), Possible Red Giant Branch star (84), Variable Star (13)\\
{\bf Molecular abs. bands} & 1274 & Variable Star (4), Long-period variable star (3), Star in Cluster (1), S Star (1), Possible Red supergiant star (1) & Infra-Red source (500), Variable Star (9), Long-period variable star (3), Star in Cluster (2), Possible Red supergiant star (2)\\
{\bf Binary stars} & 1817 & Rotationally variable Star (2), Eclipsing binary of Algol type (detached) (1), Star in Cluster (1), Double or multiple star (1), Spectroscopic binary (1) & Infra-Red source (230), Rotationally variable Star (6), Double or multiple star (4), Spectroscopic binary (3), Variable Star (2)\\
{\bf H$\alpha$/H$\beta$ emission} & 215 & High proper-motion Star (5), Rotationally variable Star (4), X-ray source (2), Double or multiple star (1), Infra-Red source (1) & Infra-Red source (33), X-ray source (10), High proper-motion Star (7), Rotationally variable Star (5), Variable Star (5)\\
{\bf Problematic$^1$} & 19095 & Star in Cluster (218), Red Giant Branch star (47), High proper-motion Star (11), Variable Star (8), Variable Star of RR Lyr type (7) & Infra-Red source (1313), Star in Cluster (253), Red Giant Branch star (48), Variable Star (17), Possible Red Giant Branch star (12)\\
\hline
\multicolumn{4}{l}{$^1$ A large fraction of such spectra are recoverable (see text).}\\
\end{tabularx}
  
\end{table*}

\def\arraystretch{1.0}
\begin{table}
\caption{List of spectra plotted in Figures \ref{hot}$-$\ref{li} representing distinct classification categories. The columns give Galah ID (unique star identifier) where available, coordinates, APASS $V$ magnitude where available, figure number, and sequential spectrum number (bottom to top) in panels for each figure. \label{specs}}
\scriptsize
\centering
\begin{tabularx}{\linewidth}{ lllccr }
\hline\hline
Gal. ID & RA (J2000) & dec (J2000) & $V$ & fig. & sp.\\
& (hh mm ss.s) & (dd mm ss.s) & (mag) & &\\
\hline
2048539 & 12h 40m 59.46s  & -45$\degr$ 27$\arcmin$ 14.1$\arcsec$ &  12.2 & \ref{hot} & 1\\
1860013 & 10h 59m 01.3s  & -47$\degr$ 34$\arcmin$ 19.9$\arcsec$ &  & \ref{hot} & 2\\
2009968 & 12h 59m 40.48s  & -45$\degr$ 53$\arcmin$ 04.4$\arcsec$ &  & \ref{hot} & 3\\
3627010 & 09h 00m 57.77s  & -27$\degr$ 09$\arcmin$ 39.8$\arcsec$ &  & \ref{hot} & 4\\
\hline
3253503 & 12h 09m 53.99s  & -31$\degr$ 25$\arcmin$ 10.5$\arcsec$ &  & \ref{cool} & 1\\
2020691 & 13h 01m 19.27s  & -45$\degr$ 45$\arcmin$ 51.8$\arcsec$ &  12.6 & \ref{cool} & 2\\
9514457 & 00h 24m 59.034s  & -72$\degr$ 07$\arcmin$ 48.33$\arcsec$ &  & \ref{cool} & 3\\
\hline
3611818 & 09h 03m 05.5s  & -27$\degr$ 20$\arcmin$ 36.9$\arcsec$ &  12.9 & \ref{molecular} & 1\\
1581751 & 17h 58m 33.44s  & -50$\degr$ 51$\arcmin$ 15.5$\arcsec$ &  & \ref{molecular} & 2\\
4935389 & 08h 07m 58.28s  & -10$\degr$ 29$\arcmin$ 30.7$\arcsec$ &  10.5 & \ref{molecular} & 3\\
1715870 & 13h 23m 49.52s  & -49$\degr$ 14$\arcmin$ 21.1$\arcsec$ &  & \ref{molecular} & 4\\
\hline
157469 & 09h 43m 55.26s  & -76$\degr$ 28$\arcmin$ 55.5$\arcsec$ &  13.4 & \ref{bin} & 1\\
1345018 & 18h 39m 12.45s  & -53$\degr$ 58$\arcmin$ 07.6$\arcsec$ &  13.3 & \ref{bin} & 2\\
1281061 & 06h 13m 09.26s  & -54$\degr$ 49$\arcmin$ 54.4$\arcsec$ &  12.6 & \ref{bin} & 3\\
2584208 & 11h 34m 04.8s  & -39$\degr$ 17$\arcmin$ 39.5$\arcsec$ &  12.7 & \ref{bin} & 4\\
3217487 & 06h 30m 58.37s  & -31$\degr$ 49$\arcmin$ 29.7$\arcsec$ &  13.6 & \ref{bin} & 5\\
2061842 & 14h 05m 59.68s  & -45$\degr$ 18$\arcmin$ 20.7$\arcsec$ &  & \ref{bin} & 6\\
\hline
6122038 & 21h 28m 56.42s  & +06$\degr$ 06$\arcmin$ 25.3$\arcsec$ &  12.4 & \ref{hahb} & 1\\
203377 & 02h 04m 32.8s  & -74$\degr$ 55$\arcmin$ 28.6$\arcsec$ &  13.2 & \ref{hahb} & 2\\
1720551 & 12h 20m 52.77s  & -49$\degr$ 11$\arcmin$ 05.7$\arcsec$ &  & \ref{hahb} & 3\\
1692074 & 12h 43m 04.53s  & -49$\degr$ 31$\arcmin$ 11$\arcsec$ &  13.1 & \ref{hahb} & 4\\
2292604 & 10h 25m 20.92s  & -42$\degr$ 41$\arcmin$ 53.9$\arcsec$ &  12.7 & \ref{hahb} & 5\\
\hline
1237165 & 07h 41m 05.11s  & -55$\degr$ 26$\arcmin$ 32.54$\arcsec$ &  13.8 & \ref{problem} & 1\\
2076959 & 19h 42m 39.47s  & -45$\degr$ 08$\arcmin$ 14.4$\arcsec$ &  13.9 & \ref{problem} & 2\\
364936 & 09h 56m 53.13s  & -70$\degr$ 57$\arcmin$ 57.8$\arcsec$ &  & \ref{problem} & 3\\
598623 & 22h 58m 39.82s  & -67$\degr$ 09$\arcmin$ 45.3$\arcsec$ &  12.8 & \ref{problem} & 4\\
226575 & 01h 13m 02.51s  & -74$\degr$ 14$\arcmin$ 22.9$\arcsec$ &  13.8 & \ref{problem} & 5\\
2791332 & 12h 41m 51.89s  & -36$\degr$ 50$\arcmin$ 37.3$\arcsec$ &  & \ref{problem} & 6\\
9520401 & 18h 23m 16.32s  & -34$\degr$ 01$\arcmin$ 27.5$\arcsec$ &  & \ref{problem} & 7\\
\hline
1230979 & 07h 37m 19.419s  & -55$\degr$ 31$\arcmin$ 44.49$\arcsec$ &  14.5 & \ref{hahb2} & 1\\
3039593 & 20h 57m 51.85s  & -33$\degr$ 52$\arcmin$ 38.4$\arcsec$ &  13.8 & \ref{hahb2} & 2\\
2400420 & 20h 08m 37.69s  & -41$\degr$ 26$\arcmin$ 45.8$\arcsec$ &  & \ref{hahb2} & 3\\
3082604 & 11h 14m 59.85s  & -33$\degr$ 22$\arcmin$ 27.8$\arcsec$ &  13.7 & \ref{hahb2} & 4\\
2414299 & 15h 24m 42.98s  & -41$\degr$ 17$\arcmin$ 09.9$\arcsec$ &  & \ref{hahb2} & 5\\
2105243 & 13h 04m 09s  & -44$\degr$ 49$\arcmin$ 18.7$\arcsec$ &  14.2 & \ref{hahb2} & 6\\
2292604 & 10h 25m 20.92s  & -42$\degr$ 41$\arcmin$ 53.9$\arcsec$ &  12.7 & \ref{hahb2} & 7\\
\hline
2504307 & 21h 28m 30.21s  & -40$\degr$ 14$\arcmin$ 30.5$\arcsec$ &  13.9 & \ref{pcyg} & 1\\
1043258 & 07h 50m 47.825s  & -58$\degr$ 43$\arcmin$ 17.31$\arcsec$ &  13.9 & \ref{pcyg} & 2\\
9518625 & 07h 45m 32.447s  & -58$\degr$ 50$\arcmin$ 26.46$\arcsec$ &  & \ref{pcyg} & 3\\
1264994 & 07h 35m 59.296s  & -55$\degr$ 03$\arcmin$ 09.68$\arcsec$ &  14.0 & \ref{pcyg} & 4\\
9519860 & 18h 05m 49.37s  & -31$\degr$ 44$\arcmin$ 26.88$\arcsec$ &  & \ref{pcyg} & 5\\
9518173 & 07h 33m 39.016s  & -55$\degr$ 34$\arcmin$ 54.83$\arcsec$ &  14.7 & \ref{pcyg} & 6\\
9518474 & 07h 41m 48.343s  & -55$\degr$ 24$\arcmin$ 54.13$\arcsec$ &  & \ref{pcyg} & 7\\
9518210 & 07h 34m 49.249s  & -56$\degr$ 02$\arcmin$ 27.53$\arcsec$ &  15.0 & \ref{pcyg} & 8\\
9518390 & 07h 39m 37.67s  & -54$\degr$ 47$\arcmin$ 05.14$\arcsec$ &  14.6 & \ref{pcyg} & 9\\
9518511 & 07h 42m 50.476s  & -55$\degr$ 26$\arcmin$ 22.95$\arcsec$ &  14.9 & \ref{pcyg} & 10\\
9519416 & 14h 13m 35.84s  & +07$\degr$ 15$\arcmin$ 08.1$\arcsec$ &  16.1 & \ref{pcyg} & 11\\
9519827 & 18h 05m 16.78s  & -31$\degr$ 38$\arcmin$ 24.72$\arcsec$ &  & \ref{pcyg} & 12\\
9519903 & 18h 06m 34.01s  & -32$\degr$ 02$\arcmin$ 07.08$\arcsec$ &  & \ref{pcyg} & 13\\
9519743 & 18h 03m 45.79s  & -32$\degr$ 20$\arcmin$ 08.52$\arcsec$ &  & \ref{pcyg} & 14\\
9517171 & 06h 43m 53.4s  & +00$\degr$ 11$\arcmin$ 10.34$\arcsec$ &  14.5 & \ref{pcyg} & 15\\
\hline
3316259 & 06h 42m 36.03s  & -30$\degr$ 42$\arcmin$ 56.1$\arcsec$ &  12.0 & \ref{li} & 1\\
1311297 & 04h 42m 27.14s  & -54$\degr$ 25$\arcmin$ 10.5$\arcsec$ &  12.5 & \ref{li} & 2\\
1260144 & 06h 22m 39.33s  & -55$\degr$ 07$\arcmin$ 13.1$\arcsec$ &  13.2 & \ref{li} & 3\\
2532175 & 15h 38m 51.73s  & -39$\degr$ 54$\arcmin$ 46.3$\arcsec$ &  13.5 & \ref{li} & 4\\
3305889 & 12h 09m 56.1s  & -30$\degr$ 50$\arcmin$ 05.8$\arcsec$ &  13.0 & \ref{li} & 5\\
\hline
\end{tabularx}
  
\end{table}

\section{Specific search for young/active stars} \label{young}

	We also present additional classification results based on a more specific projection map, in contrast to the general one presented in Section \ref{classes}. These results follow the same procedure as explained in Section \ref{classp}, but with different t--SNE input parameters and input spectral ranges. The motivation for this approach is to search for stars in their early phases of evolution \citep{2013ApJ...776..127Z} for which features in H$\alpha$, H$\beta$, and $^7$Li spectral lines can be diagnostic of their activity \citep{2010ARAA..48..581S,2014EAS....65..289J}. Perplexity is set to 50 and the spectral ranges $4841-4881$ {\AA} (H$\beta$) and $6543-6583$ {\AA} (H$\alpha$) are selected for the first t--SNE projection of the whole working set, while perplexity of 15 and reduced spectral ranges ($4859-4863, 6561-6565, 6706-6710$ {\AA}) around the three diagnostic lines  are selected for the second t--SNE projection of the filtered working set. Other combinations of perplexity and spectral ranges were tried, but this one produced the most useful projection map.
	
\begin{figure}[!htp]
  		\centering
  		\includegraphics[trim = 10mm 14mm 15mm 25mm, clip, width=1\linewidth]{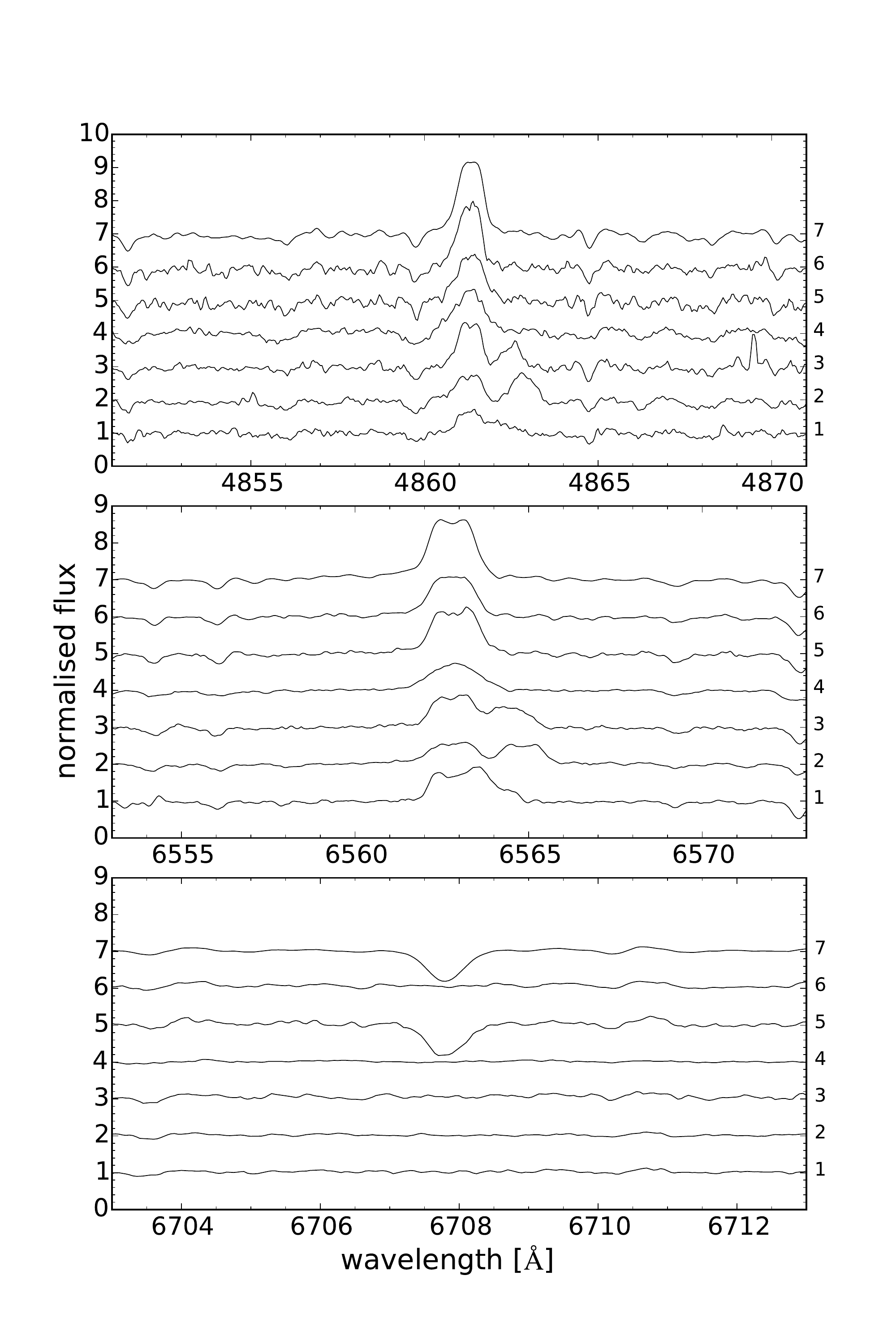}  		
		\caption{Same as Figure \ref{hot} but based on results from the specific search for young/active stars. Examples for \textit{H$\alpha$/H$\beta$ emission} category are displayed.}
		\label{hahb2}
	\end{figure}

\begin{figure}[!htp]
  		\centering
  		\includegraphics[trim = 10mm 24mm 13mm 35mm, clip, width=1\linewidth]{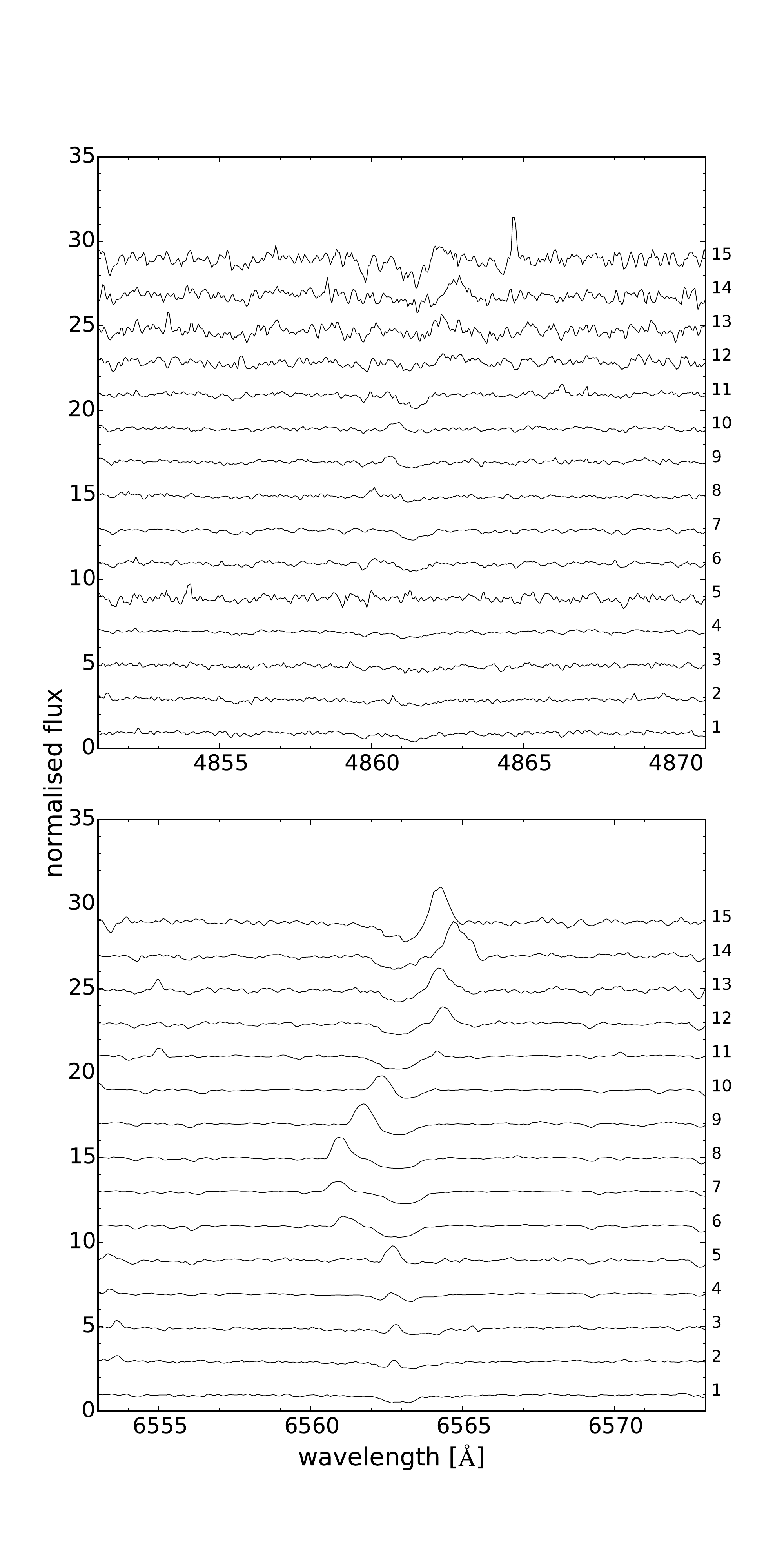}  		
		\caption{Same as Figure \ref{hahb2} but for the categories \textit{H$\alpha$/H$\beta$ emission superimposed on absorption}, \textit{H$\alpha$/H$\beta$ inverted P-Cygni}, and \textit{H$\alpha$/H$\beta$ P-Cygni}. From bottom to top in panels, each category features five examples of spectra.}
		\label{pcyg}
	\end{figure}

	Compared to the general classification from the previous section, we find additional candidates in the categories of \textit{Binary stars} (522), \textit{Problematic} spectra with oscillating continuum (665), and \textit{H$\alpha$/H$\beta$ emission} (868). With this projection map, we are able to partition the latter category and identify four distinct morphological subtypes (see Figures \ref{hahb2} and \ref{pcyg}): \textit{H$\alpha$/H$\beta$ emission}, \textit{H$\alpha$/H$\beta$ emission superimposed on absorption}, \textit{H$\alpha$/H$\beta$ P-Cygni}, and \textit{H$\alpha$/H$\beta$ inverted P-Cygni}, all indicative of diverse underlying physical processes (\citealp{2015AA...581A..52T} and references therein). The \textit{H$\alpha$/H$\beta$ emission} category is a counterpart to the one presented in the previous section, and contains diverse multicomponent profiles of H$\alpha$/H$\beta$ emission lines, that were not clearly separated in the projection map, as in the case of the latter three categories. It is possible that some emission profiles are a consequence of reduction issues instead of intrinsic properties of stars and their environment. We plan to address this possibility in future classification studies. 
	
\begin{figure}[!htp]
  		\centering
  		\includegraphics[trim = 10mm 18mm 10mm 25mm, clip, width=1\linewidth]{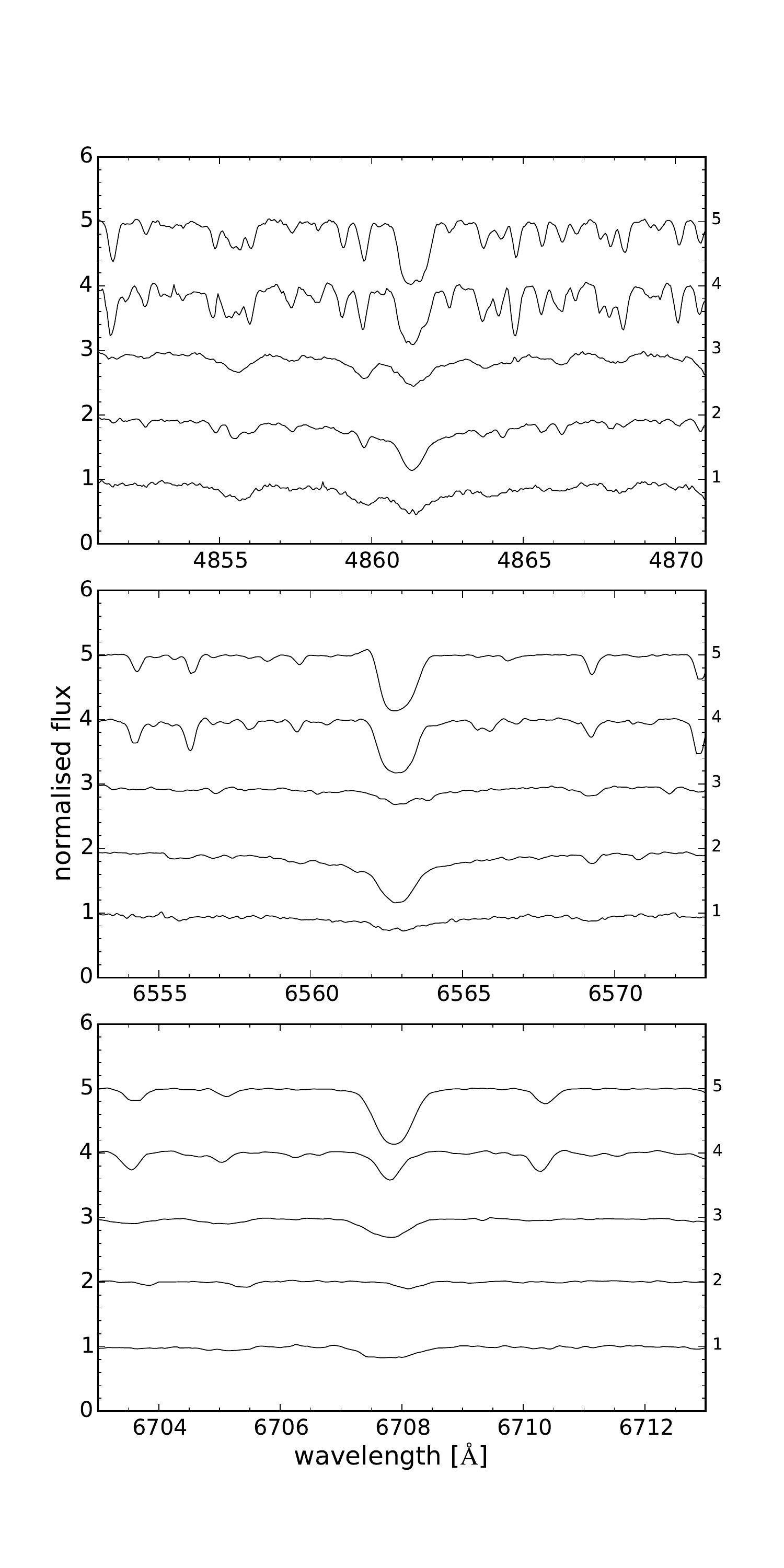}  		
		\caption{Same as Figure \ref{hahb2} but for the \textit{Lithium absorption} category.}
		\label{li}
	\end{figure}

	A new category \textit{Lithium absorption} is defined to account for spectra which display varying equivalent widths of the $^7$Li line, from weak to very strong absorptions, as shown in Figure \ref{li}. Significant $^7$Li absorption sometimes accompanies spectra in the H$\alpha$/H$\beta$ emission categories as evident from Figure \ref{hahb2}.
	
	The categories in this section are listed in Table \ref{simbad2}, along with the SIMBAD classes which indicate the connection between the youth and activity of stars and the observed H$\alpha$/H$\beta$ multicomponent profiles and prominent lithium absorption.

\def\arraystretch{1.7}
\begin{table*}
\caption{Same as Table \ref{simbad} but for the specific projection map produced in the search for young/active stars (see Section \ref{young}). The categories {\it binary stars}, {\it H$\alpha$/H$\beta$ emission}, and {\it problematic} are already defined in Section \ref{classes}, while the others are described in Section \ref{young}. \label{simbad2}}
\footnotesize
\centering
\begin{tabularx}{\linewidth}{ XrXX }
\hline\hline
Category & N & Main type & Other types\\
\hline
{\bf Binary stars} & 1428 & Star in Cluster (5), Variable Star of RR Lyr type (3), Eclipsing binary of Algol type (detached) (2), Variable Star (1), Rotationally variable Star (1) & Infra-Red source (178), Variable Star (6), Star in Cluster (5), Rotationally variable Star (4), Variable Star of RR Lyr type (3)\\
{\bf H$\alpha$/H$\beta$ emission} & 135 & Rotationally variable Star (2), High proper-motion Star (1), Double or multiple star (1), X-ray source (1), Infra-Red source (1) & Infra-Red source (19), X-ray source (7), Variable Star (4), Rotationally variable Star (3), High proper-motion Star (3)\\
{\bf H$\alpha$/H$\beta$ emission superimposed on absorption} & 479 & Star in Cluster (5) & Infra-Red source (8), Star in Cluster (5)\\
{\bf H$\alpha$/H$\beta$ P-cygni} & 18 & Variable Star (1) & Variable Star (1), Rotationally variable Star (1), Infra-Red source (1)\\
{\bf H$\alpha$/H$\beta$ inv. P-cygni} & 345 &  & \\
{\bf Lithium absorption} & 664 & Red Giant Branch star (6), Rotationally variable Star (5), High proper-motion Star (1), Spectroscopic binary (1), Pre-main sequence Star (1) & Infra-Red source (173), X-ray source (8), Rotationally variable Star (6), Red Giant Branch star (6), Pre-main sequence Star (6)\\
{\bf Problematic$^1$} & 1902 &  & Infra-Red source (174)\\
\hline
\multicolumn{4}{l}{$^1$ A large fraction of such spectra are recoverable (see text).}\\
\end{tabularx}
  
\end{table*}

\section{Catalogue} \label{catalogue}

The final classification results are collected in the catalogue, whose contents are described in Table \ref{cattab}. Spectra with at least one assigned category from either Section \ref{classes} or \ref{young} are listed by their catalogue ID, internal Galah ID of the corresponding target, coordinates of the target, APASS \citep{2012JAVSO..40..430H,2014JAD....20....4M} $V$ magnitude, classification category, and supplementary information from SIMBAD, VizieR, OGLE, and ADS databases. 

We cross-matched the coordinates of stars to retrieve information from the SIMBAD, VizieR, OGLE, and ADS on-line databases. Epoch 2000.0 coordinates of our targets are not identical with those from the catalogues, so we adopt a search radius of one arcsec where applicable. The results of the search in VizieR catalogues are retrieved in the wavelength ranges: Gamma--ray, X--ray, EUV, UV, Optical, IR, and Radio. In the catalogue, we list the number of VizieR tables in which a match is found. We also state the type of variability (class) for matched targets in OGLE-III on-line Catalogue of Variable Stars \citep{2013AcA....63...21S}. References from the literature (ADS) should provide additional information about objects of interest, but are not necessarily reliable sources of the characteristics of a certain object. 

Some of the 28,579 stars (with 31,050 spectra) in our catalogue are known to be peculiar types and are already discussed in the literature or listed in different sources. SIMBAD matched 5956 targets, VizieR finds at least one match in at least one of its catalogues for all unique targets (28,579), OGLE matched 148 targets, and 350 targets are matched successfully with references from the ADS database.

The electronic version of the catalogue will be made publicly available at the CDS, excluding results for spectra from the \textit{Problematic} category, since these mainly stand out for data reduction reasons and will be recoverable in the upgraded versions of the reduction pipeline.

\def\tabularxcolumn#1{m{#1}}
\def\arraystretch{1.2}
\begin{table*}
\caption{Description of the content for the catalogue containing results of our classification (see Section \ref{catalogue}). The full table will be available at the CDS.\label{cattab} } 
\begin{tabularx}{\linewidth}{ l l X }
\hline\hline
Label & Unit & Description\\
\hline
Catalogue\_ID & & \\
Galah\_ID & & Unique star identifier\\
DATEOBS & & Date and time of the observation\\
RA & $\degr$ & RA (J2000)\\
DEC & $\degr$ & DEC (J2000)\\
Class\_cat\_general & & General classification category as given in Section \ref{classes}\\
Class\_cat\_specific & & Specific classification category as given in Section \ref{young}\\
SIMBAD\_main\_id & & Main ID of the source in SIMBAD\\
SIMBAD\_angular\_distance & $\arcsec$ & Angular distance of Galah target to the source in SIMBAD\\
SIMBAD\_main\_type & & SIMBAD \textit{main type}\\
SIMBAD\_other\_types & & SIMBAD \textit{other types}\\
VizieR\_n\_Radio & & Number of VizieR tables for the Radio wavelength range in which Galah target has a match\\
VizieR\_n\_IR & & As VizieR\_n\_Radio but for the IR wavelength range\\
VizieR\_n\_optical & & As VizieR\_n\_Radio but for the optical wavelength range\\
VizieR\_n\_UV & & As VizieR\_n\_Radio but for the UV wavelength range\\
VizieR\_n\_EUV & & As VizieR\_n\_Radio but for the EUV wavelength range\\
VizieR\_n\_Xray & & As VizieR\_n\_Radio but for the X-ray wavelength range\\
VizieR\_n\_Gammaray & & As VizieR\_n\_Radio but for the Gamma-ray wavelength range\\
OGLE\_class & & OGLE variable star type (class)\\
ADS\_literature & & A comma-separated list of articles (title and bibcode)\\
\hline
\end{tabularx}

\end{table*}

\section{Visualisation - t--SNE Explorer} \label{visu}
	The t--SNE or \textit{Galah Explorer} is an interactive web application developed for members of the Galah collaboration that provides a visualisation of the ``feature'' based distribution of spectra in the t--SNE projection map. The basic view contains:
	\begin{description}
		\item[\bf t--SNE map] similar to those in Figures \ref{step12} and \ref{step3}. The map is split into hexagons which are colour coded based on average values of parameters of contained datapoints. 
		\item[\bf Large hexagonal frame] that displays datapoints of a selected hexagon from the map, where each datapoint is colour coded depending on selected numerical or descriptive (e.g. classification) parameters. 
		\item[\bf List of parameters] which can be chosen for colour coding the map, while the individual values for all parameters are always displayed for the currently selected datapoint. Any supplementary information on the corresponding object (star) can also be displayed together with a link to Simbad and Vizier matches.
		\item[\bf Plotting area] with four panels corresponding to the four Galah spectral bands, where the median and the dispersion of normalised fluxes of all spectra of the currently selected hexagon is displayed, over-plotted with the currently selected datapoint (spectrum).
		\item[\bf Search fields] where the user can search by Galah identifier or other parameter values or labels available in the Galah database of reduced spectra. This immediately selects and displays the matching object.
	\end{description}	
	
	The presented segments of the \textit{Galah Explorer} enable the user to locate specific areas of interest in the map that feature characteristic values of parameters. It is also possible to search for a specific object using e.g. its unique identifier. Once selected, the user can inspect its morphological vicinity (parent hexagon) with the neighbouring spectra, using statistical plots to evaluate their similarity.
	
	The (briefly) described functionality offers a very powerful and useful way of reviewing any kind of a dataset, locating and exploring its inherent structure in feature space and detecting outliers. Many different projection maps can be incorporated into the \textit{Galah Explorer} together with different DBSCAN modes for an efficient selection and identification of distinct morphological collections of spectra. The tool will be available on the Galah official website at \url{http://galah-survey.org}.

\section{Discussion} \label{disc}

We have demonstrated that t--SNE can be used as an efficient tool for discovery of diverse spectral features and classification of stellar spectra. By projecting the Galah dataset onto a two--dimensional space it is able to preserve and visually reveal its complex morphological structure. Although not tested on our sample of spectra, it was shown by \cite{citeulike:3749741} that t--SNE is far superior in its domain, putting emphasis on (1) modeling dissimilar datapoints by means of large pairwise distances, and (2) modeling similar datapoints by means of small pairwise distances, which is not obvious for other non--linear dimensionality reduction techniques and even less so for the linear ones. 

The complexity of spectral morphologies in principle increases with increasing wavelength range. In this respect, Galah, with its four spectral bands, surpasses many other spectroscopic surveys. Consequently, this makes the task of classification more difficult because spectral features can appear differently and from different effects in each band (wavelength-dependent markers of physical processes, reduction issues, hardware malfunction, different optical paths, etc.). 

Our classification procedure can accept any arbitrary spectral ranges selected by the user, which (1) enables an emphasis on the particular physics we are interested in, and (2) removes possibly unwanted influences from strong features in other parts of the spectrum, which can complicate classification of the desired types of objects. These advantages were demonstrated with the specific projection in Section \ref{young}, selecting only narrow regions of H$\alpha$, H$\beta$, and $^7$Li for the search of young, active stars. Many detections of such objects were not possible with the first projection as the full spectral information along with strong problematic features, e.g. in the IR band, concealed those in other bands. For the same reason, we might miss some interesting morphological categories with weaker characteristic features hidden by stronger ones. The specific projection map yielded new candidates for three already defined categories from the general map, while also providing four new categories, validating the principle of exploiting different t--SNE set--ups to select the best (or several) projection maps for classification purposes.

The search for peculiar objects and the classification presented here is not exhaustive or absolutely representative of the whole dataset. It is limited by and reflects our choice of the t-SNE and DBSCAN algorithms that form the basis of our analysis, the selection of their parameters, our iterative approach, and the spectral range used in each of the steps of our classification scheme. In this respect, we recognise only the most prominent features revealed by the selected classification set--up, which further define 10 distinct categories listed in Tables \ref{simbad} and \ref{simbad2}, containing a total of 31,050 spectra (28,579 unique targets). We acknowledge the possibility of establishing additional categories of exotic spectra in the Galah dataset, and this will be explored in future studies.

Using more than one projection map, possibly produced by different input selected spectral ranges, or simply using different DBSCAN modes, it can happen that the same spectrum is assigned to more than one category, due to the previously discussed reasons. Additional factor contributing to such cases are morphologically similar features, such as double lines from binary stars and emission superimposed on absorption. These might be located close in the projection space, with a possible overlap region. Spectra with more than one category can be easily identified in the catalogue (Table \ref{cattab}) as having values for both general and specific classification fields.  

The novel dimensionality reduction technique t--SNE is capable of representing astronomical spectra in a low dimensional space where their morphology and hidden features can be efficiently discovered and studied. This was shown with an effective classification of the largest astronomical high-resoluton spectroscopic dataset so far, comprising 209,533 spectra with each containing 13,600 values of flux. All data products along with the t--SNE Explorer will be publicly available in the coming data releases. The source code is freely available on--line, and our custom procedure for classification is easy to adapt to different spectroscopic or other astronomical datasets. This work will facilitate further investigation and understanding of the still--growing Galah dataset and enable focused studies of distinct categories of objects (e.g. binary stars).

Although this work has made use of external sources, it is not dependent on them and they serve mostly to support this proof of concept for classification of a wide variety of astronomical data.

\acknowledgments
Based on data acquired through the Australian Astronomical Observatory, via programs 2013B/13, 2014A/25, 2015A/19. SLM acknowledges the support of the Australian Research Council through grant DE140100598. This research has made use of the SIMBAD database \citep{2000AAS..143....9W} and of the VizieR catalogue access tool operated at CDS, Strasbourg, France. The original description of the VizieR service was published in \cite{2000AAS..143...23O}.

\bibliography{sample}

\begin{thebibliography}{}
\expandafter\ifx\csname natexlab\endcsname\relax\def\natexlab#1{#1}\fi

\bibitem[{{Asplund} {et~al.}(in prep.)}]{asplund}
{Asplund}, M., {et~al.} in prep., \mnras

\bibitem[{{Bailer-Jones} {et~al.}(1998){Bailer-Jones}, {Irwin}, \& {von
  Hippel}}]{1998MNRAS.298..361B}
{Bailer-Jones}, C.~A.~L., {Irwin}, M., \& {von Hippel}, T. 1998, \mnras, 298,
  361

\bibitem[{{Carretta} {et~al.}(2009){Carretta}, {Bragaglia}, {Gratton},
  {Lucatello}, {Catanzaro}, {Leone}, {Bellazzini}, {Claudi}, {D'Orazi},
  {Momany}, {Ortolani}, {Pancino}, {Piotto}, {Recio-Blanco}, \&
  {Sabbi}}]{2009AA...505..117C}
{Carretta}, E., {Bragaglia}, A., {Gratton}, R.~G., {et~al.} 2009, \aap, 505,
  117

\bibitem[{{Dalton} {et~al.}(2012){Dalton}, {Trager}, {Abrams}, {Carter},
  {Bonifacio}, {Aguerri}, {MacIntosh}, {Evans}, {Lewis}, {Navarro}, {Agocs},
  {Dee}, {Rousset}, {Tosh}, {Middleton}, {Pragt}, {Terrett}, {Brock}, {Benn},
  {Verheijen}, {Cano Infantes}, {Bevil}, {Steele}, {Mottram}, {Bates},
  {Gribbin}, {Rey}, {Rodriguez}, {Delgado}, {Guinouard}, {Walton}, {Irwin},
  {Jagourel}, {Stuik}, {Gerlofsma}, {Roelfsma}, {Skillen}, {Ridings},
  {Balcells}, {Daban}, {Gouvret}, {Venema}, \& {Girard}}]{2012SPIE.8446E..0PD}
{Dalton}, G., {Trager}, S.~C., {Abrams}, D.~C., {et~al.} 2012, in \procspie,
  Vol. 8446, Ground-based and Airborne Instrumentation for Astronomy IV, 84460P

\bibitem[{{Daniel} {et~al.}(2011){Daniel}, {Connolly}, {Schneider},
  {Vanderplas}, \& {Xiong}}]{2011AJ....142..203D}
{Daniel}, S.~F., {Connolly}, A., {Schneider}, J., {Vanderplas}, J., \& {Xiong},
  L. 2011, \aj, 142, 203

\bibitem[{{de Jong} {et~al.}(2012){de Jong}, {Bellido-Tirado}, {Chiappini},
  {Depagne}, {Haynes}, {Johl}, {Schnurr}, {Schwope}, {Walcher}, {Dionies},
  {Haynes}, {Kelz}, {Kitaura}, {Lamer}, {Minchev}, {M{\"u}ller}, {Nuza},
  {Olaya}, {Piffl}, {Popow}, {Steinmetz}, {Ural}, {Williams}, {Winkler},
  {Wisotzki}, {Ansorge}, {Banerji}, {Gonzalez Solares}, {Irwin}, {Kennicutt},
  {King}, {McMahon}, {Koposov}, {Parry}, {Sun}, {Walton}, {Finger}, {Iwert},
  {Krumpe}, {Lizon}, {Vincenzo}, {Amans}, {Bonifacio}, {Cohen}, {Francois},
  {Jagourel}, {Mignot}, {Royer}, {Sartoretti}, {Bender}, {Grupp}, {Hess},
  {Lang-Bardl}, {Muschielok}, {B{\"o}hringer}, {Boller}, {Bongiorno}, {Brusa},
  {Dwelly}, {Merloni}, {Nandra}, {Salvato}, {Pragt}, {Navarro}, {Gerlofsma},
  {Roelfsema}, {Dalton}, {Middleton}, {Tosh}, {Boeche}, {Caffau}, {Christlieb},
  {Grebel}, {Hansen}, {Koch}, {Ludwig}, {Quirrenbach}, {Sbordone}, {Seifert},
  {Thimm}, {Trifonov}, {Helmi}, {Trager}, {Feltzing}, {Korn}, \&
  {Boland}}]{2012SPIE.8446E..0TD}
{de Jong}, R.~S., {Bellido-Tirado}, O., {Chiappini}, C., {et~al.} 2012, in
  \procspie, Vol. 8446, Ground-based and Airborne Instrumentation for Astronomy
  IV, 84460T

\bibitem[{{De Silva} {et~al.}(2007){De Silva}, {Freeman}, {Asplund},
  {Bland-Hawthorn}, {Bessell}, \& {Collet}}]{2007AJ....133.1161D}
{De Silva}, G.~M., {Freeman}, K.~C., {Asplund}, M., {et~al.} 2007, \aj, 133,
  1161

\bibitem[{{De Silva} {et~al.}(2015){De Silva}, {Freeman}, {Bland-Hawthorn},
  {Martell}, {de Boer}, {Asplund}, {Keller}, {Sharma}, {Zucker}, {Zwitter},
  {Anguiano}, {Bacigalupo}, {Bayliss}, {Beavis}, {Bergemann}, {Campbell},
  {Cannon}, {Carollo}, {Casagrande}, {Casey}, {Da Costa}, {D'Orazi}, {Dotter},
  {Duong}, {Heger}, {Ireland}, {Kafle}, {Kos}, {Lattanzio}, {Lewis}, {Lin},
  {Lind}, {Munari}, {Nataf}, {O'Toole}, {Parker}, {Reid}, {Schlesinger},
  {Sheinis}, {Simpson}, {Stello}, {Ting}, {Traven}, {Watson}, {Wittenmyer},
  {Yong}, \& {{\v Z}erjal}}]{2015MNRAS.449.2604D}
{De Silva}, G.~M., {Freeman}, K.~C., {Bland-Hawthorn}, J., {et~al.} 2015,
  \mnras, 449, 2604

\bibitem[{Ester {et~al.}(1996)Ester, Kriegel, Jorg, \& Xu}]{citeulike:3509601}
Ester, M., Kriegel, H.-p., Jorg, S., \& Xu, X. 1996, in Proceedings of 2nd
  International Conference on KDD, 226--231

\bibitem[{{Freeman} \& {Bland-Hawthorn}(2002)}]{2002ARAA..40..487F}
{Freeman}, K., \& {Bland-Hawthorn}, J. 2002, \araa, 40, 487

\bibitem[{{Freeman}(2012)}]{2012ASPC..458..393F}
{Freeman}, K.~C. 2012, in Astronomical Society of the Pacific Conference
  Series, Vol. 458, Galactic Archaeology: Near-Field Cosmology and the
  Formation of the Milky Way, ed. W.~{Aoki}, M.~{Ishigaki}, T.~{Suda},
  T.~{Tsujimoto}, \& N.~{Arimoto}, 393

\bibitem[{{Gilmore} {et~al.}(2012)}]{2012Msngr.147...25G}
{Gilmore}, G., {et~al.} 2012, The Messenger, 147, 25

\bibitem[{{Gratton} {et~al.}(2012){Gratton}, {Carretta}, \&
  {Bragaglia}}]{2012AARv..20...50G}
{Gratton}, R.~G., {Carretta}, E., \& {Bragaglia}, A. 2012, \aapr, 20, 50

\bibitem[{{Gulati} {et~al.}(1994){Gulati}, {Gupta}, {Gothoskar}, \&
  {Khobragade}}]{1994ApJ...426..340G}
{Gulati}, R.~K., {Gupta}, R., {Gothoskar}, P., \& {Khobragade}, S. 1994, \apj,
  426, 340

\bibitem[{Hartigan \& Wong(1979)}]{hartigan1979algorithm}
Hartigan, J.~A., \& Wong, M.~A. 1979, Journal of the Royal Statistical Society.
  Series C (Applied Statistics), 28, 100

\bibitem[{{Henden} {et~al.}(2012){Henden}, {Levine}, {Terrell}, {Smith}, \&
  {Welch}}]{2012JAVSO..40..430H}
{Henden}, A.~A., {Levine}, S.~E., {Terrell}, D., {Smith}, T.~C., \& {Welch}, D.
  2012, Journal of the American Association of Variable Star Observers
  (JAAVSO), 40, 430

\bibitem[{Hinton \& Roweis(2002)}]{hinton2002sne}
Hinton, G., \& Roweis, S. 2002, in Advances in Neural Information Processing
  Systems 15 (MIT Press), 833--840

\bibitem[{Hotelling(1936)}]{hotelling1936relations}
Hotelling, H. 1936, Biometrika, 28, 321

\bibitem[{{Ibata} \& {Irwin}(1997)}]{1997AJ....113.1865I}
{Ibata}, R.~A., \& {Irwin}, M.~J. 1997, \aj, 113, 1865

\bibitem[{Izenman(2008)}]{Izenman2008}
Izenman, A.~J. 2008, Linear Discriminant Analysis (New York, NY: Springer New
  York), 237--280

\bibitem[{{Jeffries}(2014)}]{2014EAS....65..289J}
{Jeffries}, R.~D. 2014, in EAS Publications Series, Vol.~65, EAS Publications
  Series, 289--325

\bibitem[{{Kos} {et~al.}(2016){Kos}, {Lin}, {Zwitter}, {{\v Z}erjal}, {Sharma},
  {Bland-Hawthorn}, {Asplund}, {Casey}, {De Silva}, {Freeman}, {Martell},
  {Simpson}, {Schlesinger}, {Zucker}, {Anguiano}, {Bacigalupo}, {Bedding},
  {Betters}, {Da Costa}, {Duong}, {Hyde}, {Ireland}, {Kafle}, {Leon-Saval},
  {Lewis}, {Munari}, {Nataf}, {Stello}, {Tinney}, {Traven}, {Watson}, \&
  {Wittenmyer}}]{2016MNRAS.tmp.1183K}
{Kos}, J., {Lin}, J., {Zwitter}, T., {et~al.} 2016, \mnras, arXiv:1608.04391

\bibitem[{{Liu} {et~al.}(2016{\natexlab{a}}){Liu}, {Asplund}, {Yong},
  {Mel{\'e}ndez}, {Ram{\'{\i}}rez}, {Karakas}, {Carlos}, \&
  {Marino}}]{2016MNRAS.tmp.1165L}
{Liu}, F., {Asplund}, M., {Yong}, D., {et~al.} 2016{\natexlab{a}}, \mnras,
  arXiv:1608.03788

\bibitem[{{Liu} {et~al.}(2016{\natexlab{b}}){Liu}, {Yong}, {Asplund},
  {Ram{\'{\i}}rez}, \& {Mel{\'e}ndez}}]{2016MNRAS.457.3934L}
{Liu}, F., {Yong}, D., {Asplund}, M., {Ram{\'{\i}}rez}, I., \& {Mel{\'e}ndez},
  J. 2016{\natexlab{b}}, \mnras, 457, 3934

\bibitem[{{Lochner} {et~al.}(2016){Lochner}, {McEwen}, {Peiris}, {Lahav}, \&
  {Winter}}]{2016ApJS..225...31L}
{Lochner}, M., {McEwen}, J.~D., {Peiris}, H.~V., {Lahav}, O., \& {Winter},
  M.~K. 2016, \apjs, 225, 31

\bibitem[{{Luo} {et~al.}(2015){Luo}, {Zhao}, {Zhao}, {Deng}, {Liu}, {Jing},
  {Wang}, {Zhang}, {Shi}, {Cui}, {Chu}, {Li}, {Bai}, {Wu}, {Cai}, {Cao}, {Cao},
  {Carlin}, {Chen}, {Chen}, {Chen}, {Chen}, {Chen}, {Chen}, {Chen},
  {Christlieb}, {Chu}, {Cui}, {Dong}, {Du}, {Fan}, {Feng}, {Fu}, {Gao}, {Gong},
  {Gu}, {Guo}, {Han}, {He}, {Hou}, {Hou}, {Hou}, {Hu}, {Hu}, {Hu}, {Huo},
  {Jia}, {Jiang}, {Jiang}, {Jiang}, {Jin}, {Kong}, {Kong}, {Lei}, {Li}, {Li},
  {Li}, {Li}, {Li}, {Li}, {Li}, {Li}, {Li}, {Li}, {Li}, {Li}, {Liang}, {Lin},
  {Liu}, {Liu}, {Liu}, {Liu}, {Lu}, {Luo}, {Mao}, {Newberg}, {Ni}, {Qi}, {Qi},
  {Shen}, {Shi}, {Song}, {Song}, {Su}, {Su}, {Tang}, {Tao}, {Tian}, {Wang},
  {Wang}, {Wang}, {Wang}, {Wang}, {Wang}, {Wang}, {Wang}, {Wang}, {Wang},
  {Wang}, {Wang}, {Wang}, {Wang}, {Wang}, {Wang}, {Wang}, {Wang}, {Wang},
  {Wang}, {Wei}, {Wei}, {Wu}, {Wu}, {Wu}, {Wu}, {Xing}, {Xu}, {Xu}, {Xu},
  {Yan}, {Yang}, {Yang}, {Yang}, {Yang}, {Yao}, {Yu}, {Yuan}, {Yuan}, {Yuan},
  {Yuan}, {Zhai}, {Zhang}, {Zhang}, {Zhang}, {Zhang}, {Zhang}, {Zhang},
  {Zhang}, {Zhang}, {Zhao}, {Zhou}, {Zhou}, {Zhu}, {Zhu}, {Zou}, \&
  {Zuo}}]{2015RAA....15.1095L}
{Luo}, A.-L., {Zhao}, Y.-H., {Zhao}, G., {et~al.} 2015, Research in Astronomy
  and Astrophysics, 15, 1095

\bibitem[{{Majewski} {et~al.}(2015){Majewski}, {Schiavon}, {Frinchaboy},
  {Allende Prieto}, {Barkhouser}, {Bizyaev}, {Blank}, {Brunner}, {Burton},
  {Carrera}, {Chojnowski}, {Cunha}, {Epstein}, {Fitzgerald}, {Garcia Perez},
  {Hearty}, {Henderson}, {Holtzman}, {Johnson}, {Lam}, {Lawler}, {Maseman},
  {Meszaros}, {Nelson}, {Coung Nguyen}, {Nidever}, {Pinsonneault}, {Shetrone},
  {Smee}, {Smith}, {Stolberg}, {Skrutskie}, {Walker}, {Wilson}, {Zasowski},
  {Anders}, {Basu}, {Beland}, {Blanton}, {Bovy}, {Brownstein}, {Carlberg},
  {Chaplin}, {Chiappini}, {Eisenstein}, {Elsworth}, {Feuillet}, {Fleming},
  {Galbraith-Frew}, {Garcia}, {Anibal Garcia-Hernandez}, {Gillespie},
  {Girardi}, {Gunn}, {Hasselquist}, {Hayden}, {Hekker}, {Ivans}, {Kinemuchi},
  {Klaene}, {Mahadevan}, {Mathur}, {Mosser}, {Muna}, {Munn}, {Nichol},
  {O'Connell}, {Robin}, {Rocha-Pinto}, {Schultheis}, {Serenelli}, {Shane},
  {Silva Aguirre}, {Sobeck}, {Thompson}, {Troup}, {Weinberg}, \&
  {Zamora}}]{2015arXiv150905420M}
{Majewski}, S.~R., {Schiavon}, R.~P., {Frinchaboy}, P.~M., {et~al.} 2015, ArXiv
  e-prints, arXiv:1509.05420

\bibitem[{Marin {et~al.}(2005)Marin, Mengersen, \& Robert}]{marin2005bayesian}
Marin, J.-M., Mengersen, K., \& Robert, C.~P. 2005, Handbook of statistics, 25,
  459

\bibitem[{{Martell} {et~al.}(2016){Martell}, {Sharma}, {Buder}, {Duong},
  {Schlesinger}, {Simpson}, {Lind}, {Ness}, {Marshall}, {Asplund},
  {Bland-Hawthorn}, {Casey}, {De Silva}, {Freeman}, {Kos}, {Lin}, {Zucker},
  {Zwitter}, {Anguiano}, {Bacigalupo}, {Carollo}, {Casagrande}, {Da Costa},
  {Horner}, {Huber}, {Hyde}, {Kafle}, {Lewis}, {Nataf}, {Stello}, {Tinney},
  {Watson}, \& {Wittenmyer}}]{2016arXiv160902822M}
{Martell}, S., {Sharma}, S., {Buder}, S., {et~al.} 2016, ArXiv e-prints,
  arXiv:1609.02822

\bibitem[{{Mason} {et~al.}(2001){Mason}, {Wycoff}, {Hartkopf}, {Douglass}, \&
  {Worley}}]{2001AJ....122.3466M}
{Mason}, B.~D., {Wycoff}, G.~L., {Hartkopf}, W.~I., {Douglass}, G.~G., \&
  {Worley}, C.~E. 2001, \aj, 122, 3466

\bibitem[{{Matijevi{\v c}}(2016)}]{2016IAUS..317..336M}
{Matijevi{\v c}}, G. 2016, in IAU Symposium, Vol. 317, The General Assembly of
  Galaxy Halos: Structure, Origin and Evolution, ed. A.~{Bragaglia},
  M.~{Arnaboldi}, M.~{Rejkuba}, \& D.~{Romano}, 336--337

\bibitem[{{Matijevi{\v c}} {et~al.}(2010){Matijevi{\v c}}, {Zwitter}, {Munari},
  {Bienaym{\'e}}, {Binney}, {Bland-Hawthorn}, {Boeche}, {Campbell}, {Freeman},
  {Gibson}, {Gilmore}, {Grebel}, {Helmi}, {Navarro}, {Parker}, {Seabroke},
  {Siebert}, {Siviero}, {Steinmetz}, {Watson}, {Williams}, \&
  {Wyse}}]{2010AJ....140..184M}
{Matijevi{\v c}}, G., {Zwitter}, T., {Munari}, U., {et~al.} 2010, \aj, 140, 184

\bibitem[{{Matijevi{\v c}} {et~al.}(2012){Matijevi{\v c}}, {Zwitter},
  {Bienaym{\'e}}, {Bland-Hawthorn}, {Boeche}, {Freeman}, {Gibson}, {Gilmore},
  {Grebel}, {Helmi}, {Munari}, {Navarro}, {Parker}, {Reid}, {Seabroke},
  {Siebert}, {Siviero}, {Steinmetz}, {Watson}, {Williams}, \&
  {Wyse}}]{2012ApJS..200...14M}
{Matijevi{\v c}}, G., {Zwitter}, T., {Bienaym{\'e}}, O., {et~al.} 2012, \apjs,
  200, 14

\bibitem[{{McGurk} {et~al.}(2010){McGurk}, {Kimball}, \&
  {Ivezi{\'c}}}]{2010AJ....139.1261M}
{McGurk}, R.~C., {Kimball}, A.~E., \& {Ivezi{\'c}}, {\v Z}. 2010, \aj, 139,
  1261

\bibitem[{{Munari} {et~al.}(2014){Munari}, {Henden}, {Frigo}, \&
  {Dallaporta}}]{2014JAD....20....4M}
{Munari}, U., {Henden}, A., {Frigo}, A., \& {Dallaporta}, S. 2014, Journal of
  Astronomical Data, 20

\bibitem[{{Ness} {et~al.}(2015){Ness}, {Hogg}, {Rix}, {Ho}, \&
  {Zasowski}}]{2015ApJ...808...16N}
{Ness}, M., {Hogg}, D.~W., {Rix}, H.-W., {Ho}, A.~Y.~Q., \& {Zasowski}, G.
  2015, \apj, 808, 16

\bibitem[{{Ochsenbein} {et~al.}(2000){Ochsenbein}, {Bauer}, \&
  {Marcout}}]{2000AAS..143...23O}
{Ochsenbein}, F., {Bauer}, P., \& {Marcout}, J. 2000, \aaps, 143, 23

\bibitem[{Pezzotti {et~al.}(2016)Pezzotti, Lelieveldt, van~der Maaten,
  H{\"o}llt, Eisemann, \& Vilanova}]{bib:pezzotti:2016}
Pezzotti, N., Lelieveldt, B., van~der Maaten, L., {et~al.} 2016, IEEE
  Transactions on Visualization and Computer Graphics

\bibitem[{{Piskunov} \& {Valenti}(2016)}]{2016arXiv160606073P}
{Piskunov}, N., \& {Valenti}, J.~A. 2016, ArXiv e-prints, arXiv:1606.06073

\bibitem[{{Pourbaix} {et~al.}(2004){Pourbaix}, {Tokovinin}, {Batten}, {Fekel},
  {Hartkopf}, {Levato}, {Morrell}, {Torres}, \& {Udry}}]{2004AA...424..727P}
{Pourbaix}, D., {Tokovinin}, A.~A., {Batten}, A.~H., {et~al.} 2004, \aap, 424,
  727

\bibitem[{{Prusti}(2012)}]{2012AN....333..453P}
{Prusti}, T. 2012, Astronomische Nachrichten, 333, 453

\bibitem[{{Sharma} \& {Johnston}(2009)}]{2009ApJ...703.1061S}
{Sharma}, S., \& {Johnston}, K.~V. 2009, \apj, 703, 1061

\bibitem[{{Sheinis} {et~al.}(2015){Sheinis}, {Anguiano}, {Asplund},
  {Bacigalupo}, {Barden}, {Birchall}, {Bland-Hawthorn}, {Brzeski}, {Cannon},
  {Carollo}, {Case}, {Casey}, {Churilov}, {Warrick}, {Dean}, {De Silva},
  {D'Orazi}, {Duong}, {Farrell}, {Fiegert}, {Freeman}, {Gabriella}, {Gers},
  {Goodwin}, {Gray}, {Green}, {Heald}, {Heijmans}, {Ireland}, {Jones}, {Kafle},
  {Keller}, {Klauser}, {Kondrat}, {Kos}, {Lawrence}, {Lee}, {Mali}, {Martell},
  {Mathews}, {Mayfield}, {Miziarski}, {Muller}, {Pai}, {Patterson}, {Penny},
  {Orr}, {Schlesinger}, {Sharma}, {Shortridge}, {Simpson}, {Smedley}, {Smith},
  {Stafford}, {Staszak}, {Vuong}, {Waller}, {de Boer}, {Xavier}, {Zheng},
  {Zhelem}, {Zucker}, \& {Zwitter}}]{2015JATIS...1c5002S}
{Sheinis}, A., {Anguiano}, B., {Asplund}, M., {et~al.} 2015, Journal of
  Astronomical Telescopes, Instruments, and Systems, 1, 035002

\bibitem[{{Soderblom}(2010)}]{2010ARAA..48..581S}
{Soderblom}, D.~R. 2010, \araa, 48, 581

\bibitem[{{Soszy{\'n}ski} {et~al.}(2013){Soszy{\'n}ski}, {Udalski},
  {Szyma{\'n}ski}, {Kubiak}, {Pietrzy{\'n}ski}, {Wyrzykowski}, {Ulaczyk},
  {Poleski}, {Koz{\l}owski}, {Pietrukowicz}, \&
  {Skowron}}]{2013AcA....63...21S}
{Soszy{\'n}ski}, I., {Udalski}, A., {Szyma{\'n}ski}, M.~K., {et~al.} 2013,
  \actaa, 63, 21

\bibitem[{{Steinmetz} {et~al.}(2006){Steinmetz}, {Zwitter}, {Siebert},
  {Watson}, {Freeman}, {Munari}, {Campbell}, {Williams}, {Seabroke}, {Wyse},
  {Parker}, {Bienaym{\'e}}, {Roeser}, {Gibson}, {Gilmore}, {Grebel}, {Helmi},
  {Navarro}, {Burton}, {Cass}, {Dawe}, {Fiegert}, {Hartley}, {Russell},
  {Saunders}, {Enke}, {Bailin}, {Binney}, {Bland-Hawthorn}, {Boeche}, {Dehnen},
  {Eisenstein}, {Evans}, {Fiorucci}, {Fulbright}, {Gerhard}, {Jauregi}, {Kelz},
  {Mijovi{\'c}}, {Minchev}, {Parmentier}, {Pe{\~n}arrubia}, {Quillen}, {Read},
  {Ruchti}, {Scholz}, {Siviero}, {Smith}, {Sordo}, {Veltz}, {Vidrih}, {von
  Berlepsch}, {Boyle}, \& {Schilbach}}]{2006AJ....132.1645S}
{Steinmetz}, M., {Zwitter}, T., {Siebert}, A., {et~al.} 2006, \aj, 132, 1645

\bibitem[{Switzer \& Green(1984)}]{switzer1984min}
Switzer, P., \& Green, A.~A. 1984, Computer science and statistics, 13

\bibitem[{{Ting} {et~al.}(2012){Ting}, {Freeman}, {Kobayashi}, {De Silva}, \&
  {Bland-Hawthorn}}]{2012MNRAS.421.1231T}
{Ting}, Y.-S., {Freeman}, K.~C., {Kobayashi}, C., {De Silva}, G.~M., \&
  {Bland-Hawthorn}, J. 2012, \mnras, 421, 1231

\bibitem[{{Traven} {et~al.}(2015){Traven}, {Zwitter}, {Van Eck}, {Klutsch},
  {Bonito}, {Lanzafame}, {Alfaro}, {Bayo}, {Bragaglia}, {Costado}, {Damiani},
  {Flaccomio}, {Frasca}, {Hourihane}, {Jimenez-Esteban}, {Lardo}, {Morbidelli},
  {Pancino}, {Prisinzano}, {Sacco}, \& {Worley}}]{2015AA...581A..52T}
{Traven}, G., {Zwitter}, T., {Van Eck}, S., {et~al.} 2015, \aap, 581, A52

\bibitem[{{{\v Z}erjal} {et~al.}(2013){{\v Z}erjal}, {Zwitter}, {Matijevi{\v
  c}}, {Strassmeier}, {Bienaym{\'e}}, {Bland-Hawthorn}, {Boeche}, {Freeman},
  {Grebel}, {Kordopatis}, {Munari}, {Navarro}, {Parker}, {Reid}, {Seabroke},
  {Siviero}, {Steinmetz}, \& {Wyse}}]{2013ApJ...776..127Z}
{{\v Z}erjal}, M., {Zwitter}, T., {Matijevi{\v c}}, G., {et~al.} 2013, \apj,
  776, 127

\bibitem[{{Valenti} \& {Piskunov}(1996)}]{1996AAS..118..595V}
{Valenti}, J.~A., \& {Piskunov}, N. 1996, \aaps, 118, 595

\bibitem[{{Valentini} {et~al.}(2016){Valentini}, {Chiappini}, {Davies},
  {Elsworth}, {Mosser}, {Lund}, {Miglio}, {Chaplin}, {Rodrigues}, {Boeche},
  {Steinmetz}, {Matijevic}, {Kordopatis}, {Bland-Hawthorn}, {Munari},
  {Bienayme}, {Gibson}, {Gilmore}, {Grebel}, {Helmi}, {Kunder}, {McMillan},
  {Navarro}, {Parker}, {Reid}, {Seabroke}, {Siviero}, {Watson}, {Wise}, \&
  {Zwitter}}]{2016arXiv160903826V}
{Valentini}, M., {Chiappini}, C., {Davies}, G.~R., {et~al.} 2016, ArXiv
  e-prints, arXiv:1609.03826

\bibitem[{{van der Maaten}(2013)}]{2013arXiv1301.3342V}
{van der Maaten}, L. 2013, ArXiv e-prints, arXiv:1301.3342

\bibitem[{van~der Maaten \& Hinton(2008)}]{citeulike:3749741}
van~der Maaten, L., \& Hinton, G. 2008, Journal of Machine Learning Research,
  9, 2579

\bibitem[{{von Hippel} {et~al.}(1994){von Hippel}, {Storrie-Lombardi},
  {Storrie-Lombardi}, \& {Irwin}}]{1994MNRAS.269...97V}
{von Hippel}, T., {Storrie-Lombardi}, L.~J., {Storrie-Lombardi}, M.~C., \&
  {Irwin}, M.~J. 1994, \mnras, 269, 97

\bibitem[{{Watson}(1987)}]{1987PhDT.........7W}
{Watson}, F.~G. 1987, PhD thesis, Edinburgh Univ.~(Scotland).

\bibitem[{{Wenger} {et~al.}(2000){Wenger}, {Ochsenbein}, {Egret}, {Dubois},
  {Bonnarel}, {Borde}, {Genova}, {Jasniewicz}, {Lalo{\"e}}, {Lesteven}, \&
  {Monier}}]{2000AAS..143....9W}
{Wenger}, M., {Ochsenbein}, F., {Egret}, D., {et~al.} 2000, \aaps, 143, 9

\bibitem[{Young(2013)}]{young2013multidimensional}
Young, F.~W. 2013, Multidimensional scaling: History, theory, and applications
  (Psychology Press)

\end{thebibliography}

\end{document}